\documentclass[nohyper]{JHEP3}

\usepackage{amsfonts,graphics,epsfig,amssymb}

\newcommand{\beq}{\begin{equation}}
\newcommand{\eeq}{\end{equation}}
\newcommand{\bea}{\begin{eqnarray}}
\newcommand{\eea}{\end{eqnarray}}
\newcommand{\beas}{\begin{eqnarray*}}
\newcommand{\eeas}{\end{eqnarray*}}

\def\de{\partial}
\def\tr{{\rm Tr}\,}
\def\Tr{ \hbox{\rm Tr}\,}
\def\Trnc{ \hbox{\rm Tr} \, }

\def\tr{ \hbox{\rm Tr}}
\def\tk{\widetilde{k}}

\def\SU{{\rm SU}}
\def\L{{\rm L}}
\def\R{{\rm R}}

\def\cl{{\rm cl}}

\def\U{{\rm U}}

\def\bra{\langle}\def\ket{\rangle}

\def\tE{\widetilde{E}}

\def\rbf{\mathbf{r}}

\def\Im{\hbox {\rm Im}\,}

\def\Z{\mathbb Z}
\def\1{\mathbbm{1}}
\def\N{{\cal N}}
\def\tq{{\widetilde q}}
\def\W{{\cal W}}
\def\tQ{{\widetilde Q}}

\def\p{{}^{\,\prime}}
\def\a{\alpha}

\def\tm{{\widetilde m}}

\def\tk{{\widetilde k}}

\def\tD{\widetilde{D}}
\def\td{\widetilde{d}}

\def\tnc{\widetilde{n}_c}
\def\tq{\widetilde{q}}
\def\snd{ \mbox{\tiny $ \N=2$ } }
\def\snu{ \mbox{\tiny $ \N=1$ } }
\def\LLambda{\Delta}

\title{A Coincidence Problem:\\
How to Flow from $\N =2$ SQCD to $ \N = 1 $ SQCD}
\author{Stefano Bolognesi\\ 
William I. Fine Theoretical Physics Institute, University of Minnesota, 
116 Church St. S.E., Minneapolis, MN 55455, USA\\
\email{bolognesi@physics.umn.edu}}

\abstract{We discuss, and propose a solution for, a still unresolved problem regarding the breaking from $\N=2$ super-QCD to $\N=1$ super-QCD. A mass term $W=\mu \Tr \Phi^2 / 2$ for the adjoint field, which classically does  the required breaking perfectly, quantum mechanically leads to a relevant operator that, in the infrared, makes the theory flow away from pure $\N=1$ SQCD.  To avoid this problem, we first need to extend the theory from $\SU (n_c)$ to $\U (n_c)$.  We then look for the quantum generalization of the condition $W^{\prime}(m)=0$, that is, the coincidence between a root of the derivative of the superpotential $W(\phi)$ and the mass $m$ of the quarks. There are $2n_c -n_f$ of such points in the moduli space. We suggest that with an opportune choice of superpotential, that selects  one of these coincidence vacua in the moduli space, it is possible to flow from $\N=2$ SQCD to $\N=1$ SQCD. Various arguments support this claim. In particular, we shall determine the exact location in the moduli space of these coincidence vacua and the precise factorization of the SW curve.
}


\preprint{FTPI-MINN-08/26; UMN-TH-2705/08}

\begin{document}

\section{Introduction}
\label{intro}

In this paper, we want to discuss a problem of supersymmetry breaking, from $\N=2$ to $\N=1$. An issue, which is not completely understood in the literature, is how to flow from $\N=2$ super-QCD to pure $\N=1$ super-QCD \cite{Argyres:1996eh}. 
We shall explain the problem, why it is still unresolved, and propose a solution for it.

$\N=1$ super-QCD contains the gauge vector multiplet $W_{\alpha}$ for the gauge group $\SU (n_c)$, and $n_f$ flavors of massless quarks $Q$ in the fundamental representation and $\tQ$ in the anti-fundamental representation. $\N=2$ super-QCD is an extension of the previous one. We must add a chiral adjoint multiplet $\Phi$ that together with $W_{\alpha}$ forms a $\N=2$ gauge supermultiplet. The quarks $Q$ and $\tQ^{\dagger}$ fit together to form an $\N=2$ matter hypermultiplet. An opportune superpotential $\tQ \Phi Q$  should also be added.

At the classical level, both theories have a moduli space of vacua. The $\N=2$ moduli space is divided into two distinct parts: the Coulomb branch and the Higgs branch. On the Coulomb branch, $\phi \neq 0$ and $q,\tq =0$. The rank of the gauge group is preserved. On the Higgs branch, some of the quarks $q$'s and $\tq^{\,}$'s develop an expectation value. The Higgs branches develop out of various singular sub-manifolds of the Coulomb branch where the quarks become effectively massless.  For what we are interested now, we can just consider the origin of the moduli space, $\phi=0$, $q,\tq =0$,  which is the meeting point between the Coulomb branch and the maximal Higgs branch.

The $\N=1$ moduli space is different. First of all, there is no Coulomb branch, since there is no $\phi$. Second, the Higgs branch is bigger than the one of $\N=2$. Now only the $D$-term determines the expectation value of $q$ and $\tq$. The $F$-term constraint $\tq q =0$ is now absent.

The simplest way to break $\N=2$ supersymmetry, while preserving $\N=1$, is to add a mass term in the superpotential for the adjoint chiral field: $\mu \Tr \Phi^2 /2$. The effect of this perturbation
is to lift all the Coulomb branch except from the origin with its maximal Higgs branch attached. Another effect happens as $\mu$ becomes very, very large. The potential for the meson $\tq q$, coming from the $F$-term of $\N=2$, becomes more and more shallow and vanishes in the limit $\mu \to \infty$. In this limit, the full Higgs branch of $\N=1$ is thus recovered.
\FIGURE{
\includegraphics[width=35em]{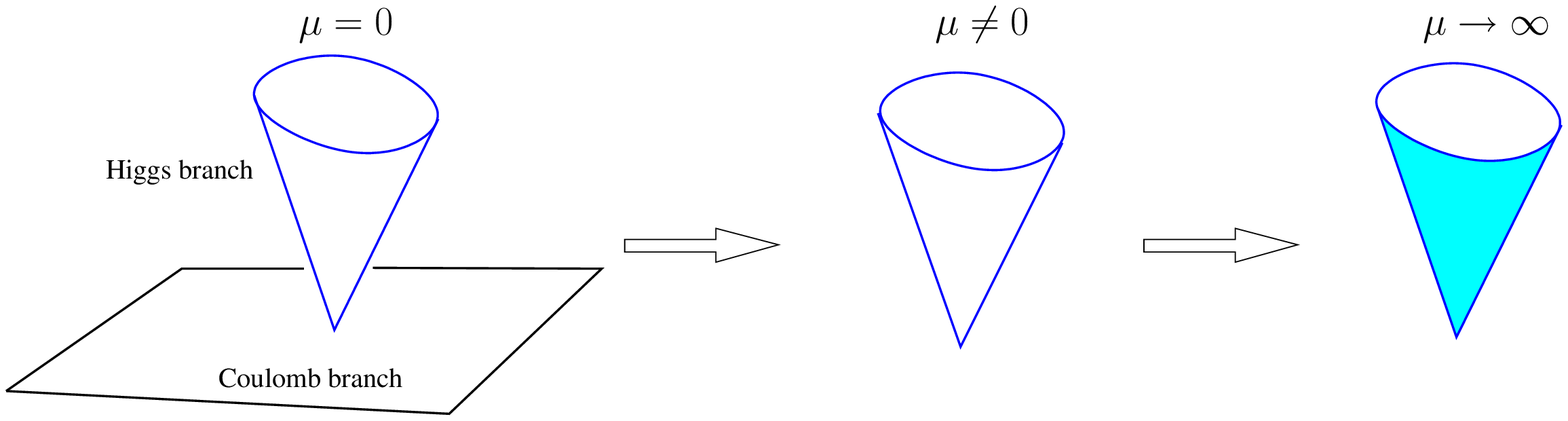}
\caption{Classical supersymmetry breaking from $\N=2$ to $\N=1$ super-QCD. The breaking is obtained by a mass parameter $\mu$ for the field $\Phi$. At $\mu =0$, we have the $\N=2$ theory that has a Coulomb branch and a maximal Higgs branch that develops from the origin of the moduli space $\phi=0$, $q$, $\tq =0$. The mass term lifts the Coulomb branch leaving only the Higgs branch. As $\mu$ goes to infinity, the Higgs branch is enhanced by other flat directions, $\tq q \neq 0$, and we recover the full moduli space of $\N=1$.
}\label{higgs}
}

\FIGURE{
\includegraphics[width=25em]{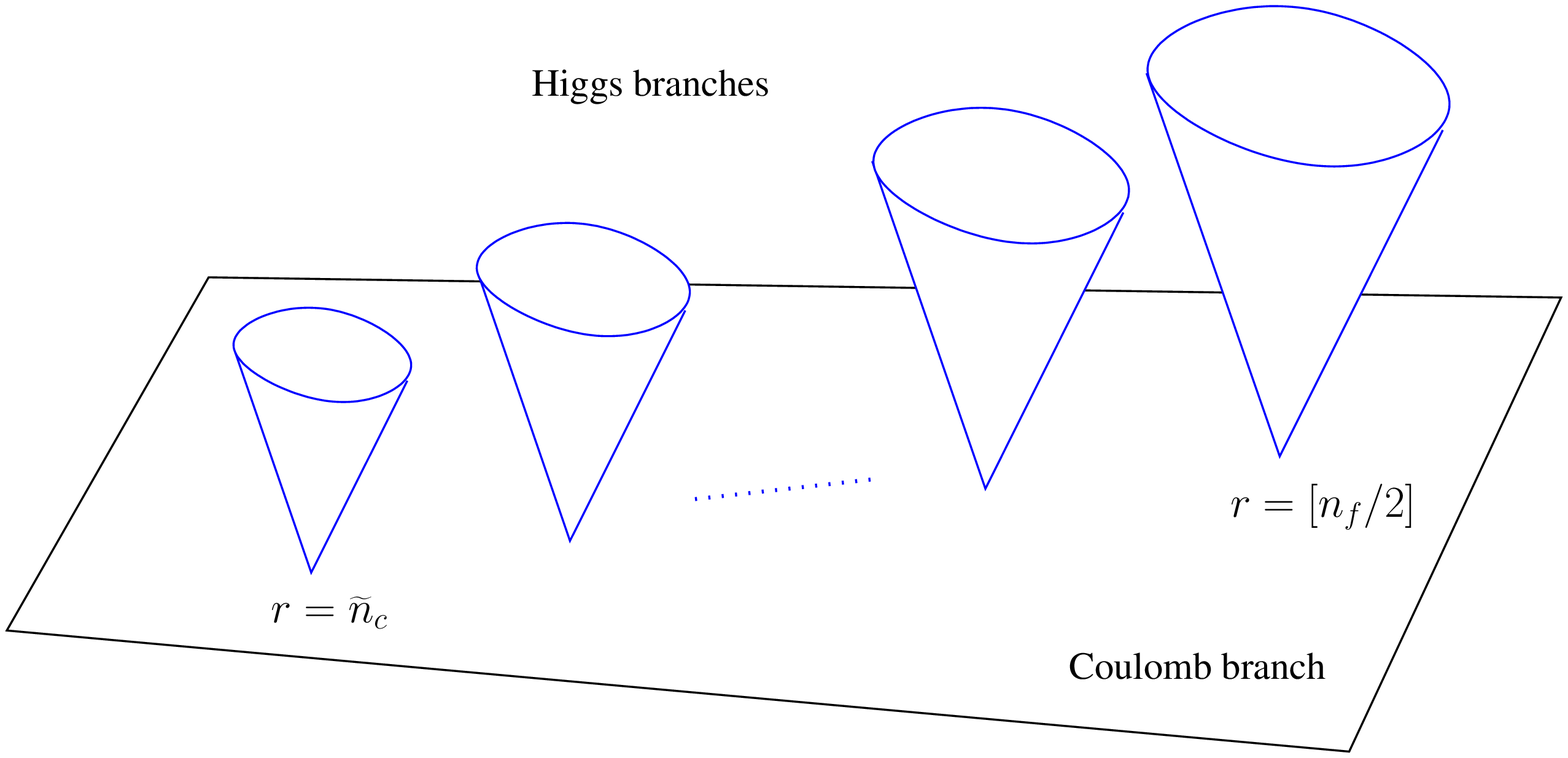}
\caption{The notion of origin of the moduli space does not exist anymore for the quantum version of $\SU (n_c)$ super-QCD. There are various points of maximal singularity labeled by an integer $r$ that runs from $\tnc=n_f -n_c$ to $[n_f /2]$. Higgs branches of increasing size (quaternionic $r(n_f -r)$) develop from these singularities.}
\label{r-vacua}
}
Quantum mechanically, it is a completely different story. The metric of the Coulomb branch of $\N=2$ is modified, and the origin of the moduli space, strictly speaking, does not exist anymore. It is instead split in many different points out of which various Higgs branches develop (Figure \ref{r-vacua}). These singularities are labeled by an integer $r$ that runs from $\tnc = n_f -n_c$ to $[n_f / 2]$.\footnote{By $[n_f/2]$, here and in the rest of the paper, we mean the integer part of $n_f/2$.}   Adding a mass perturbation $\mu \Tr \Phi^2 /2$, we lift all the Coulomb branch with the exception of these singular points and their Higgs branches. Every one of these points gives a particular $\N=1$ theory, but {\it none} of them flows to pure $\N=1$ super-QCD in the $\mu \to \infty$ limit.

The mechanism to flow from $\N=2$ super-QCD to $\N=1$ super-QCD, which is very simple in   classical theory, does not work in   quantum theory. In this paper, we shall present a proposal to make this mechanism work. This will certainly open new possibilities in the study of the $\N=1$ dynamics and its rich dualities.

\vskip 0.50cm
\begin{center}
*  *  *
\end{center} 
\noindent
The ideas we shall present originate, in part, from some recent developments in the theory of the Abelian and non-Abelian heterotic vortex-string. It is good to say a few words about this subject since it could help to understand better what follows.

$\U (n_c)$ $\N=2$ theories with quark hypermultiplets with mass $m$, can be put in a weak Higgs phase with a suitable Fayet-Iliopoulos (FI) term for the $\U (1)$ gauge field. This term does not break $\N=2$ supersymmetry; it only breaks the $\SU (2)_R$ symmetry. There is a supersymmetric vacuum where $\phi=m $,  so that the quarks are effectively massless and can develop a condensate.  At the Higgs breaking energy scale, the dynamics are essentially transferred to the $1+1$ action of the non-Abelian vortex. Being the vortex half-BPS, the $1+1$ effective action has $\N=(2,2)$ supersymmetries on its worldsheet. The tension of the vortex is proportional to the central charge $T=4 \pi \xi$, where $\xi$ is the FI term.

A superpotential for the adjoint field $\Phi$ can be added in order to break supersymmetry down to $\N=1$, for example, the single trace of an holomorphic function $\W= \Tr  W(\Phi)$. The vacuum structure is not changed, but now the string has tension $4 \pi \sqrt{\xi^2 + |W^{\prime}(m)|^2}$ and loses its BPS properties. $W^{\prime}(m)$ is the value of the superpotential evaluated at the mass of the quarks \cite{Auzzi:2004yg}.

Something very special happens when a zero of $W^{\prime}$ coincides with the mass $m$. In this case, the tension is back to the central charge $T=4 \pi \xi$, and the vortex effective theory re-acquires part of the supersymmetries. The vortex is half-BPS, and the worldsheet theory has a $\N=(2,0)$ supersymmetries inherited by the bulk $\N=1$.
Additional studies show that the strong dynamics on the $1+1$ theory dynamically break supersymmetry \cite{Shifman:2005st,Edalati:2007vk,Shifman:2008wv}.

These studies, up to now, leave an unresolved important question: \textit{is the coincidence condition $W^{\prime}(m)=0$ quantum-mechanically meaningful?} To explain  the question better, let us call $a$ one of the roots of $W^{\prime}$. Classically we can choose to fine tune the parameters so that $a$ is exactly equal to $m$. The non-trivial question is if this fine tuning is stable or not under quantum corrections. Note that both $a$ and $m$ are in general renormalized by instanton corrections. The quantum stability of this coincidence condition becomes a non-trivial question in the non-Abelian theory.

The heterotic string gives an ambiguous answer to these questions. On one hand, there is the enhancement of supersymmetry in the particular case $a=m$. Symmetry enhancement is in general what protects from quantum corrections. On the other hand, from the analysis of the strong dynamics of the worldsheet theory, we know that $\N=(2,0)$ supersymmetry is dynamically broken. But this, for the present paper, is not our concern.

\vskip 0.50cm
\begin{center}
*  *  *
\end{center} 
\noindent
Now back to the main subject of the paper, the $4d$ dynamics {\it without}  the FI term. 
The present paper originates from the question we previously asked, but for the present work, we shall restrict ourselves to the case of zero  Fayet-Iliopoulos term. We thus want to answer these questions:  \textit{1) Is the coincidence condition $W^{\prime}(m)=0$ quantum-mechanically meaningful?} \textit{2) And if it is so, how to express this coincidence condition in the quantum setup?}

We shall see that the solution of this coincidence problem  will give us the right vacuum and superpotential to flow from $\N=2$ super-QCD to $\N=1$ super-QCD.

The previous digression about the heterotic vortex-string gives us the first clue on how to solve the problem. First of all, $\SU (n_c)$ is not the right environment in which to search; we should extend the theory to $\U (n_c)$. The only price we pay is the loss of asymptotic freedom for the global $\U (1)$. But this is not a big problem. We can always think that the theory is embedded in a bigger and finite theory (for example, a bigger $\SU (N_c)$ theory) that at a certain energy scale is broken to $\U (n_c)$.

The $\N=2$ $\U (n_c)$  has, with respect to $\SU (n_c)$, one more dimension in the Coulomb branch, namely the coordinate $u_1 = \Tr \phi$. We have already said that in the Coulomb branch of $\SU (n_c)$ there is no remnant of the classical origin of the moduli space. There are instead various points labeled by an integer $r$ from $\tnc$ to $[n_f/2]$. Things are different extending the moduli space with one more dimension, $\Tr \phi$. It is in this bigger environment that we shall find the quantum analogs of the origin of the moduli space; actually we shall find $2n_c - n_f$ of them.

Our goal is now to find the quantum generalization of the coincidence points defined by $W^{\prime}(m)=0$. In other terms, when a root $a$ of $W^{\prime}$ coincides with the mass of the quarks $m$. When $a$ and $m$ are generic, we have two kinds of vacua. One is a Coulomb vacuum, with $\phi=a$, and the quark has effective mass $a-m$. The other is a Higgs vacuum where $\phi=m$ (color-flavor locking), and the quarks condense: $\tq q=W^{\prime}(m)$. In the case of coincidence, we have $\phi=m=a$. The quarks are massless and do not condense.

Quantum mechanically, the quark condensate, in a generic $r$ vacuum,  is given by\footnote{The case $r=\tnc$ is an exception to this rule. Here there is an extra massless particle, and the dual-quark can thus have zero condensate.} \cite{Bolognesi:2004da}
\beq
\label{formula}
\tq q = r \left. \sqrt{ {W^{\prime}(z)}^2 + f(z)} \right|_{z=m} \ .
\eeq
Where the terms in the square root are given by the factorization of the Seiberg-Witten curve,
\bea
\label{curveandfactorization}
{y}^2 &=& \frac{1}{4} P_{n_c}(z)^2-\Lambda^{2n_c-n_f}\prod_{i=1}^{n_f}(z-m_i) \nonumber \\
  &=&
\frac{1}{4 {g_{k}}^2}\left({W\p}(z)^2+f(z)\right)H_{n_c-k}(z)^2 \ .
\eea
$H(v)^2$ is the polynomial containing the double roots corresponding to particles that become massless. The rest is contained in ${W^{\prime}(z)}^2 +f(z)$ where $W$ is the classical superpotential and $f(z)$ is the quantum modification. The effect of $f(z)$  is to split the roots, otherwise doubled, of $W^{\prime}(z)^2$. For the vacua we are interested in, the $r$ vacua, the superpotential has degree $k=2$.
So quantum mechanically the roots of $W^{\prime}(z)$ are split in two. To find a generalization of the coincidence points, we should take one of these zeros and, moving in the $\Tr \phi$ direction, make it collide with the bunch of zeros at $z=m$ corresponding to the quark singularity (Figure \ref{rcollision}).
\FIGURE{
\includegraphics[width=25em]{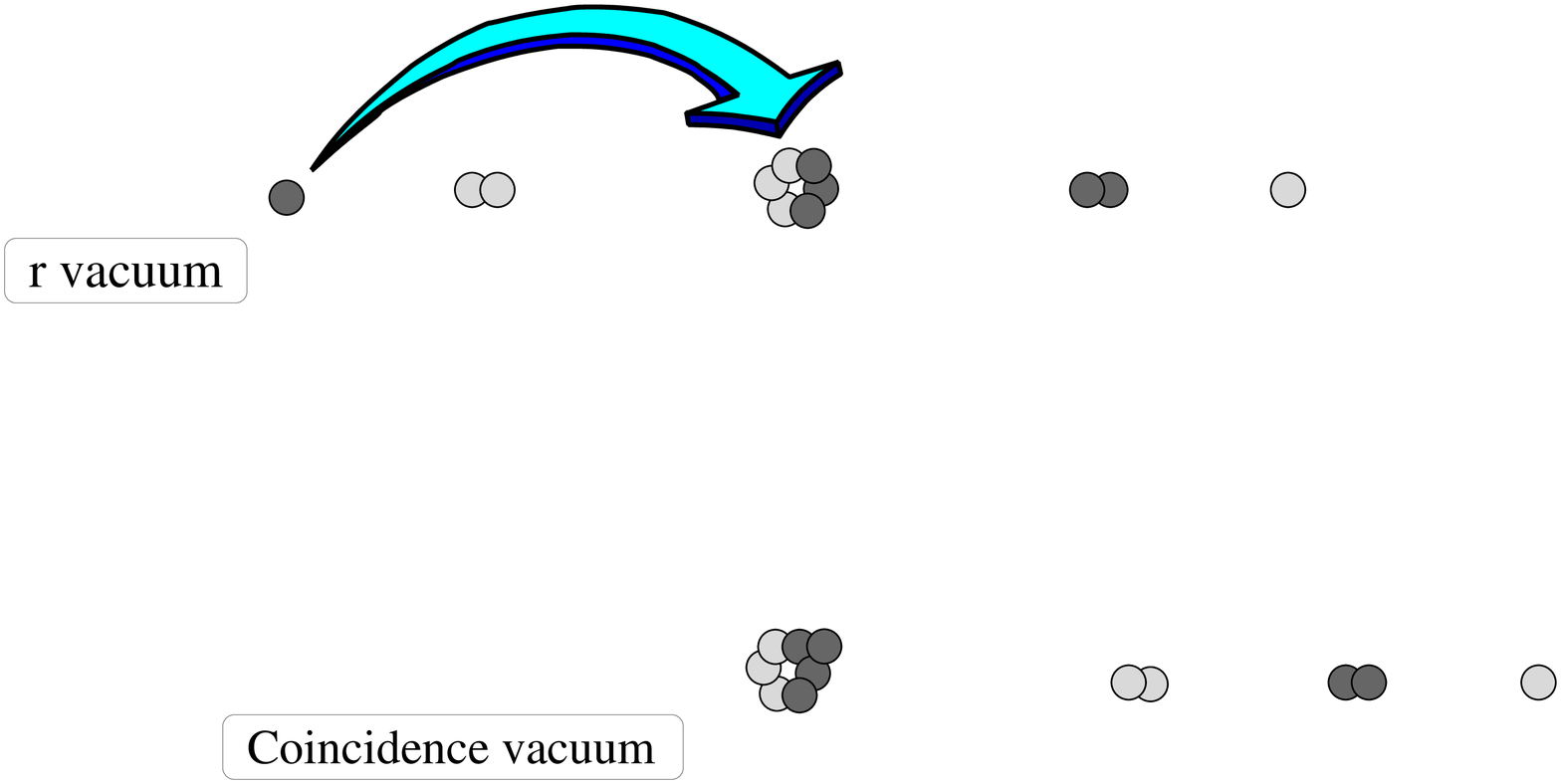}
\caption{In a generic $r$ vacuum (in the example we have the $n_c,n_f =6,6$ and $r=3$), all the roots of the Seiberg-Witten curve are paired, apart from two of them. These two roots correspond to the polynomial ${W\p}(z)^2+f(z)$ in the curve factorization. To obtain the coincidence point, we need to move one of these two unpaired roots and make it coincide with the bunch of roots at the quark mass. We can achieve this changing the value of the coordinate $\Tr \phi$ in the moduli space. In Section \ref{General}, we shall find the precise factorization of the curve in these coincidence vacua.}
\label{rcollision}
}

Doing this, we get something else for free: {\it all the $r$ vacua go to the same coincidence points}. This is one of the strongest indications that tells us that these $2n_c -n_f$ points are the quantum generalization of the origin of the moduli space in   classical theory. 
\FIGURE{
\includegraphics[width=25em]{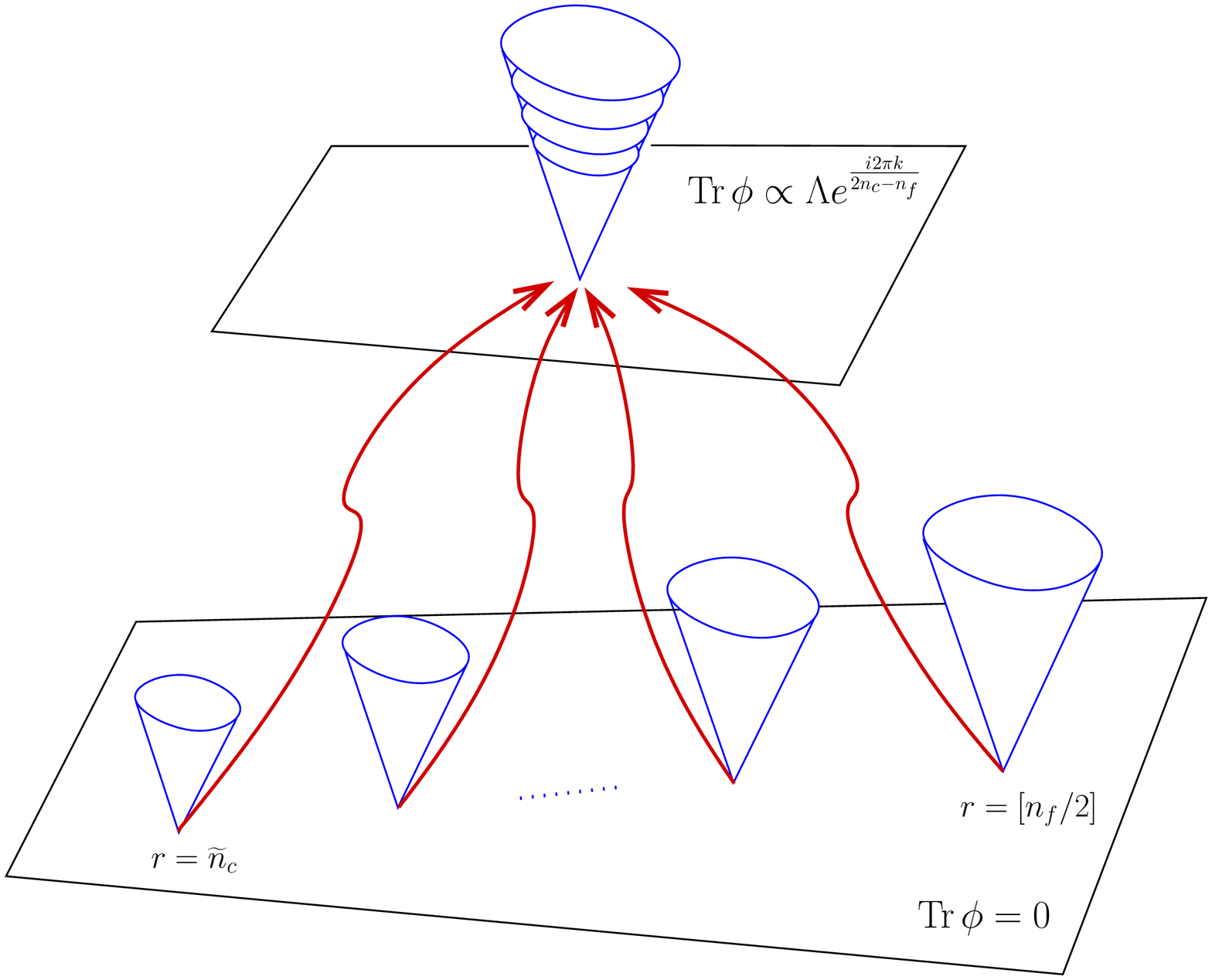}
\caption{Moving away from the $\Tr \phi=0$ section of the moduli space, it is possible to make all the $r$ vacua to collide in the coincidence points. The Higgs branch that emanates from these points is the maximal one, and all the other Higgs branches are incorporated into it. There are $2n_c - n_f$ of these points and are situated at $\Tr \phi =(2n_c -n_f) \Lambda e^{i2\pi k / (2n_c - n_f)}$.}
\label{higgs-branches-collision}
}
Every $r$ vacuum is a point in which the curve has a multiple root $z^{2r}$ corresponding to massless quarks and all the other roots, except two, are doubled. We then change the $\Tr \phi$ coordinate, keeping all the pairing of the roots, and we move until one of the unpaired roots collides with the bunch of quark zeros $z^{2r}$. We can do the same for all the $r$ vacua. They all intersect in the coincidence points, and the Higgs branches are all included in the maximal one (Figure \ref{higgs-branches-collision}).

\vskip 1.00cm

\vskip 0.50cm
\begin{center}
*  *  *
\end{center} 
\noindent
We organize the paper as follows. In Section \ref{abelian}, we discuss the simple example of super-QED. Now there is no strong dynamics in the infrared, and everything works out smoothly.  In Section \ref{nonabelian}, we introduce the non-Abelian $\U (n_c)$ theories. We provide the needed information for the rest of the paper.  In Section \ref{Example} we describe the quantum coincidence in the first non-trivial examples, starting from $2$ colors and $2$ flavors.  In Section \ref{General}, we describe the coincidence points in the general case. An explicit formula shall be found for them and for the superpotential that select them.  In Section \ref{MQCD}, we take a look from the MQCD perspective. We conclude in Section \ref{conclusion} with discussion about the renormalization group (RG) flow and summarize the arguments in favor of our claim.


\section{Warm-up with the Abelian Theory}
\label{abelian}

A good place to start, in explaining the notion of coincidence points, is the $\U(1)$ Abelian theory. We shall now introduce these three theories: $\N=2$ super-QED, broken $\N=2$ super-QED by a mass term, and pure $\N=1$ super-QED. These theories are infrared free, so there is no strong dynamics that could chance the classical vacuum structure and phases. The notion of coincidence point and its quantum stability are thus straightforward.

The action for $\N=2$ super-QED, expressed in $\N=1$ superfields notation, is a sum of the Kahler term plus a superpotential term
 \bea
\label{classicalSQED}
{\cal L}&=& \int d^2\theta d^2\bar{\theta} \, \left(\frac{1}{e^2} \Phi^\dagger \Phi+Q^\dagger_i e^V Q^i + \tQ^{  \dagger i} e^{-V}{\widetilde Q}_i \right) +\nonumber\\
&&+\int d^2\theta \, \left(\frac{1}{4e^2}W^{\a}W_{\a}+ \W(\Phi,Q,\tQ) \right)  + {\rm h.c.} \ , 
\eea
where
\beq
\W(\Phi,Q,\tQ)=\sqrt{2}\left({\widetilde Q}\Phi Q-m{\widetilde Q}_iQ^i \right) \ .
\eeq
The potential for the scalar fields is the sum of $F$ terms and the $D$ term
\beq
\label{pot}
V=2|(\phi-m)q^i|^2+2|(\phi-m)\tq_i|^2+2e^2|\tq_i q^i|^2+\frac{e^2}{2}(|q^i|^2-|\tq_i|^2)^2\ .
\eeq
The index $i$ represents the flavor and runs from $1$ to $n_f$. 
The moduli space consists of a Coulomb branch and a Higgs branch. The Coulomb branch is parameterized by the expectation value of $\phi$, and the quark condensate instead vanishes. When $\phi = m$, the effective quark mass vanishes and a Higgs branch develops from the Coulomb one.  The Higgs branch is a manifold parameterized by the following algebraic equations
\bea
\tq_i q^i &=& 0 \ , \nonumber \\
|q^i|^2-|\tq_i|^2 &=& 0 \ . 
\eea 
For $n_f = 1$ there is no Higgs branch. For greater $n_f$, it is a manifold of complex dimension $2n_f -2$. Higgs branches, in $\N=2$ theories, are conic hyper-Kahler manifolds (in $\N=1$ theories are only Kahler).

Now we break $\N=2$  to $\N=1$ by  means of the  superpotential $ W(\Phi)$ that is a holomorphic function of $\Phi$
\beq
\W(\Phi,Q,\tQ)=\sqrt{2}\left({\widetilde Q}\Phi Q-m{\widetilde Q}_iQ^i - W(\Phi) \right)  \ . 
\eeq
The potential for the scalar fields is now modified into
\beq
\label{potW}
V=2|(\phi-m)q^i|^2+2|(\phi-m)\tq_i|^2+2e^2|\tq_i q^i  -  W\p(\phi)|^2+\frac{e^2}{2}(|q^i|^2-|\tq_i|^2)^2 \ .
\eeq
There are now two kinds of vacua. One is when the value of $\phi$ is equal to some zero of the derivative of the superpotential. We denote $a_j$ these roots so that $W^{\prime}(z)=\prod_{j=1}^{k} (z-a_j)$ where $k+1$ is the degree of the superpotential. There are $k$ vacua where $\phi=a_i$. These are Coulomb vacua because the expectation values of $q$ and $\widetilde{q}$ are zero.  Then there is a Higgs vacuum where $\phi$ is locked to the hypermultiplet mass $m$. Here  $q$ and $\widetilde{q}$ develop an expectation value 
\beq \tq q={W}^{\prime}(m)  \ , \qquad  q=\tq^{\,*} \ . \eeq
The number of vacua is thus $k+1$. They all preserve supersymmetry.

Things are different when one root of $W^{\prime}$, say $a_1$, is exactly equal to the hypermultiplet mass. From here on, we shall use the word {\it coincidence} to indicate this particular circumstance.  In this coincidence case, we have only $k$ vacua instead of $k+1$.  $k-1$ of them are Coulomb vacua, and as before, they arise when $\phi=a_j$ for $j=2, \dots k$.  The last vacuum is when $\phi=m=a_1$.  It is a kind of hybrid between the Higgs and the Coulomb vacua previously described. The phase is Coulomb, because the expectation value of the quark condensate vanishes. But it is a Coulomb phase of a different nature from the previous one. In the previous case, the quark hypermultiplet was massive, with effective mass equal to $a_1 - m$.  So at low energy we had only a gauge vector multiplet (and being $\U(1)$ the gauge coupling is finite and the theory is free). In the coincidence case, the effective quark mass is $a_1-m=0$,  so at low energy the $\U(1)$ vector multiplet is coupled to a massless quark (and being $\U(1)$ the gauge coupling runs to zero in the infrared).  Now what about the quantum stability?  In principle, we can always tune the parameters of the bare Lagrangian so that we have a coincidence between the mass and a root of the superpotential.  The question is if this tuning is stable or not under quantum corrections.   Parameters in the superpotential, like the roots $a_j$ and the mass $m$, are subject to the non-renormalization theorem and thus unchanged by perturbative quantum correction. They could be modified by the  non-perturbative dynamics (that will be the case in the following part of the paper), but for the Abelian case we do not have to worry about that. This implies the quantum stability of this coincidence.

We now consider the quadratic superpotential
\beq
W(\phi) = \mu \left (\frac{\phi^2}{2} - m \phi \right) \ ,
\eeq
so that the derivative is the linear polynomial $W^{\prime}(z)= 2 \mu (z-m)$, and we choose the root to coincide exactly with the quark mass.   Integrating out the superfield $\Phi$,  we get 
\beq \Phi=m+\frac{1}{\mu} \tQ Q \ ,
\eeq 
and   the effective  superpotential
 \beq
\W(Q,\tQ)=\sqrt{2}\left( \frac{1}{2} \mu m^2 + \frac{1}{2 \mu} \tQ Q \, \tQ Q  \right)   \ .
\eeq
The potential for the scalar fields is
\beq
\label{potWmu}
V=\frac{2}{ \mu^2}| \tq q \,  q^i |^2+\frac{2}{ \mu^2}| \tq q \,  \tq_i|^2
+\frac{e^2}{2}(|q^i|^2-|\tq_i|^2)^2\ .
\eeq
With a gauge and flavor transformation, we can always bring a generic solution into the form 
\bea
q&=&(k,0,0,\dots,0) \ , \nonumber \\
\tq &=& (0,k,0, \dots ,0)\ , \qquad k \in \mathbb{R}^+ \ .
\eea
At this point, the potential for the $\tq q$ meson field is
\beq
\label{potWinfty}
V (\tq q )=\frac{4}{ \mu^2 } (\tq q )^2 k^2+ \dots \ .
\eeq
In the $\mu \to \infty$, we have an enhancement of the moduli space because the $F$ term condition $\tq q=0$ loses progressively its weight. The direction $\tq q \neq 0$, which is generically lifted by the potential (\ref{potWmu}), becomes more and more shallow and finally flat in the $\mu \to \infty$ limit. Note that at the origin of the Higgs branch (where it touches the Coulomb branch), the meson $\tq q$ is always massless, but the $\tq q$ direction is still lifted by the higher order potential quartic in the meson field.

Pure $\N=1$ 
super-QED has the following Lagrangian 
\bea
\label{classicalSQEDN=1}
{\cal L}&=& \int d^2\theta d^2\bar{\theta} \, \left( Q_i^\dagger e^V Q^i+\tQ^{\dagger i} e^{-V}{\widetilde Q}_i \right) +\nonumber\\
&&+\int d^2\theta \, \frac{1}{4e^2}W^{\a}W_{\a}  + {\rm h.c.}  \ ,
\eea
and the potential for the scalar fields is given just by the $D$ term
\beq
\label{potN=1}
V=\frac{e^2}{2}(|q^i|^2-|\tq_i|^2)^2\ .
\eeq
With a gauge and flavor transformation, we can bring a solution into the form 
\bea
&& q=(k,0,0,\dots,0) \ , \nonumber \\
&& \tq = (\tk,\lambda,0, \dots ,0)\ , \nonumber \\
&&  k,\tk,\lambda   \in \mathbb{R}^+ \ , \qquad k^2 = \tk^2 + \lambda^2 \ .
\eea

\vskip 0.10cm
\noindent
So to summarize here what happens to the moduli space for various values of $\mu$:
\begin{itemize}
\item   The $\N=2$ theory ($\mu =0$)  has a Coulomb branch with complex dimension, one parameterized by $\phi$. A Higgs branch with complex dimension $2n_f -2$  develops from the origin $\phi =0$. 
\item   Switching on the mass $\mu$, the Coulomb branch is lifted, and only the origin $\phi=0$ survives with its Higgs branch attached.
\item  In the $\mu \to \infty$ limit, the $F_{\phi}$ condition $\tq q =0$ is no more effective due to the $1/\mu$ suppression. The Higgs branch is enhanced and becomes $2n_f -1$ dimensional. We thus smoothly recover the moduli space of $\N=1$ SQED.
\end{itemize}


\section{Non-Abelian Theory}
\label{nonabelian}


The non-Abelian $\U (n_c)$ super QCD has the following action\footnote{All the time we use $\Tr$ in the paper, we mean a trace in the color space.}
\bea
\label{sqcdN=2}
{\cal L}&=& \int d^2\theta d^2\bar{\theta} \,  \frac{2}{e^2} \Trnc (\Phi^\dagger e^V \Phi e^{-V})+ \sum_{i=1}^{n_f}\, ( Q_i^\dagger e^{V}Q^i+\widetilde{Q}_i e^{-V} {\widetilde{Q}}^{\dagger i} ) +\nonumber\\
&&+\int d^2\theta \, \left(\frac{1}{2e^2} \Trnc (W^{\alpha}W_{\alpha})+ \W(\Phi,Q,\tQ) \right)  + {\rm h.c.}  \ ,
\eea
where the superpotential is
\beq
\W(\Phi,Q,\tQ)=\sum_{i=1}^{n_f} \, \sqrt{2}(\widetilde{Q}_i\Phi Q^i - m_i\widetilde{Q}_i Q^i ) \ ,
\eeq
and $m^{i}$ are the masses for the flavors with the index $i=1,\dots,n_f$.
The charges and symmetries of the fields are given in Table \ref{chargesn=2}.
\TABLE[h]{
\begin{tabular}{c|ccccccc}
& $\U (n_c) $& $\times$ & $\SU (n_f)$ & $\times $&$ \U (1)_R$ & $\times $& $\U (1)_J$ \\ \hline
$Q$                  & $\bf n_c$      && $\bf n_f $       &&$ 0 $            
&& $1$ \\
$\tQ$                & $\bf \bar n_c  $&&$ \bf \bar n_f$  && $0$            
&&$ 1$ \\
$\Phi$               & $\bf adj  $    && $\bf 1 $         &&$ 2 $            
&&$ 0$ \\
$\Lambda_{\snd}^{2n_c-n_f}$ &$ \bf 1$    && $\bf 1$     && $2(2n_c{-}n_f) $
&&$ 0$ \\
\hline
\end{tabular} \label{chargesn=2} \caption{Fields, symmetries, and charges for $\N=2$ SQCD.}
}
The gauge group is $\U (n_c) = \SU (n_c) \times \U (1) / \Z_{n_c}$.  At a very high energy scale, which we call $\Lambda_{\mbox{\tiny $\mathrm{cutoff}$}}$, the gauge couplings for the Abelian $\U (1)$ and for the non-Abelian $\SU (n_c)$ are the same.\footnote{At this scale, the theory becomes part of a bigger, asymptotically free, theory. An example could be an $\SU (n_c +1)$ gauge theory that is broken to $\U (n_c)$ at the scale $\Lambda_{\mbox{\tiny $\mathrm{cutoff}$}}$.}
Then the gauge couplings run in opposite directions. The non-Abelian running is given by
\beq
\frac{1}{e_{\mbox{\tiny $\SU (n_c)$}}^2}=(2n_c-n_f)\log{\left(\frac{\Lambda_{\snd}}{\mu}\right)} \ ,
\eeq 
and becomes strong in the infrared.
The $\U(1)_R$ symmetry is anomalous and broken to $\Z_{2n_c -n_f}$ by instanton zero modes. 
This fact, as usual, can be elegantly incorporated giving a $\U(1)_R$ charge to the dynamical scale $\Lambda_{\snd}$ as in Table \ref{chargesn=2}.

We then break supersymmetry adding a superpotential $W(\Phi)$ for the adjoint field  $\Phi$.  The superpotential term becomes
\beq
\W(\Phi,Q,\tQ)=\sum_{i=1}^{n_f} \, \sqrt{2}(\widetilde{Q}_i\Phi Q^i - m_i\widetilde{Q}_i Q^i ) - \sqrt{2} \Trnc W(\Phi) \ ,
\eeq
where
\beq
\label{tree}
W(z)=\sum_{j=0}^{k} \, \frac{g_j}{j+1}z^{j+1}\ ,\qquad W^\prime (z)=g_k\prod_{j=1}^{k}\, (z-a_j)\ .
\eeq
The adjoint scalar field $\phi$, from the $D$ term,  is a $n_c \times n_c$ complex-hermitian matrix that can thus be diagonalized. 
The eigenvalues follow the same rule explained for the Abelian case. 
In the supersymmetric vacua, the diagonal elements of the adjoint field must be equal to a flavor mass $m_i$ (the color-flavor locking case) or to a root $a_j$ of $W^\prime$. 
There can also be some degeneracy in the eigenvalues. A generic solution is
\beq
\bra\phi\ket=\left(\begin{array}{cccc}
m_i \mathbf{1}_{r_i}&&&\\
&\ddots&&\\
&&a_j {\mathbf 1}_{n_j}&\\
\end{array}\right)\ ,
\qquad
\sum_{j=1}^{k} \, n_j+\sum_{i=1}^{n_f}\, r_i=n_c\ .
\eeq
The adjoint field breaks the  gauge group to $ \prod_{i=1}^{n_f} \U(r_i) \times \prod_{j=1}^{k} \U (n_j)$. The quarks locked to the adjoint field are effectively massless and condense after the perturbation of $W(\Phi)$. The quark condensate $W\p(m_i)$ breaks the residual gauge groups $\prod_{i=1}^{n_f} \U(r_i) $. In the low-energy, we are thus left with a $\prod_{j=1}^{k} \U (n_j)$ super-Yang-Mills theory without quarks.

\vskip 0.10cm
In the case of coincidence, we have to consider also another possibility; the previous analysis is not exhaustive.  We call $h$ the number of coincidences. So we choose the last $h$ roots of $W^{\prime}$ to be equal to the last $h$ masses
\beq
s_l = a_{k-h+l} = m_{n_f - h+l} \ , \qquad l=1,\dots,h \ .
\eeq 
The other roots $a_j$ for $j=1,\dots,k-h$ and the other masses $m_i$ for $i=1,\dots,n_f-h$ are generic (no coincidence). 
Now the eigenvalues of $\phi$ can be equal to some $s_l$  with $l=1,\dots,h$, or to $a_j$ with $j=1, \dots,k-h$ or to $m_i$ with $i=1, \dots, n_f-h$.  The generic structure is thus
\beq
\bra\phi\ket=
\left(
\begin{array}{ccccc}
  s_l  \mathbf{1}_{n_l} &&&   &   \\
  & \ddots &&& \\
  &&  a_j \mathbf{1}_{n_j} & &  \\
  &&& \ddots & \\
  &&&   &   m_i \mathbf{1}_{n_i}
\end{array}
\right) \ ,
\eeq
and the gauge group is broken as
\beq
\U (n_c) \rightarrow \prod_{l=1}^{h} \U (n_l) \times \prod_{j=1}^{k-h} \U (n_j) \times \prod_{i=1}^{n_f} \U (n_i) \ .
\eeq 
The various residual gauge groups have different fate according to the three cases.  In the first case, we have at low energy $\U (n_l)$ SQCD with a certain number of \emph{massless} flavors.  The second case leaves $\U (n_j)$ pure SYM, which has $n_j$ discrete supersymmetric vacua where the gauge group is in the confinement phase. In the third case, the quarks acquire an expectation value $\widetilde{Q}Q=W^{\prime}(m_i)$. The gauge group is thus in the Higgs phase.   The fact that the first case has a completely different phase structure is an indication that the coincidence we talking about should be taken quantum mechanically seriously.

\vskip 0.10cm
We now break to $\N=1$ supersymmetry by turning on a
bare mass $\mu$ for the adjoint superfield $\Phi$.  
In the microscopic theory,
this corresponds to an $\N=1$ theory with a superpotential
\beq
{\cal W} = \sqrt2\, \left( \tQ_i \Phi Q^i - {\mu \over 2} \Trnc \Phi^2 \right).
\eeq
For $\mu\gg\Lambda_{\snd}$ we can integrate $\Phi$ out in a weak-coupling
approximation, obtaining an effective quartic superpotential
\beq
\label{operator}
{\cal W}^\prime =  \frac{1}{{\mu \sqrt{2} }}  \Tr (Q\tQ Q\tQ) \ .
\eeq
Classically, in the limit $\mu {\rightarrow} \infty$ this superpotential becomes
negligible, and we find $\N=1$ $\U (n_c)$ super-QCD with $n_f$ flavors
and no superpotential. Quantum mechanically, we cannot say that since (\ref{operator}) is a relevant operator and in general, as we shall see more in detail, makes the theory to flow away from the right coincidence point.

\vskip 0.10cm
Pure $\N=1$ SQCD has no adjoint field $\Phi$, and the Lagrangian is
\bea
{\cal L}&=& \int d^2\theta d^2\bar{\theta} \, \sum_{i=1}^{n_f}\, (Q^{\dagger}_i e^{V}Q^i+\widetilde{Q}_i e^{-V}\widetilde{Q}^{\dagger i} ) +\nonumber\\
&&+\int d^2\theta \, \frac{1}{2e^2} \Trnc (W^{\alpha}W_{\alpha}) + {\rm h.c.}  \ .
\eea
Fields and charges are given in Table \ref{chargesn=1}. 
\TABLE[h]{
\begin{tabular}{c|ccccccc}
& $\SU (n_c)$ & $\times$ & $\SU (n_f)_\L$ & $\times$ & $\SU (n_f)_\R$  & $\times$ & $\U (1)_R$ \\ \hline
$Q$                 & $\bf n_c$      && $\bf n_f$  && $\bf 1 $   && $0$              \\
$\tQ$                & $\bf \bar n_c$ && $\bf 1$ && $\bf \bar n_f$  && $0$             
\\
$\Lambda_{\snu}^{3n_c - n_f} $& $\bf 1 $       && $\bf 1$     && $\bf 1  $&& $ (3n_c -n_f)$ \\
\hline
\end{tabular}
\label{chargesn=1}
\caption{Fields, symmetries and charges for $\N=1$ SQCD.}
}
Note that now the non-Abelian flavor symmetry is enhanced to $\SU(n_f)_\L \times \SU(n_f)_\R$. Left and right quarks can be rotated independently. The running of the  non-Abelian gauge coupling, due to the absence of the field $\phi$, is now a little bit faster
\beq
\frac{1}{e_{\mbox{\tiny $  \SU (n_c)$}} ^2} = (3n_c-n_f)\log{\left(\frac{\Lambda_{\snu}}{\mu}\right)} \ .
\eeq 
The $\U(1)_R$ symmetry is anomalous and broken to $\Z_{3n_c - n_f}$.

The matching of the dynamical scales between the $\N=2$ and the $\N=1$ theory is given by
\beq
\label{relationofscales}
\Lambda_{\snu}^{3n_c-n_f} = \mu^{n_c} \Lambda_{\snd}^{2n_c-n_f}\ .
\eeq
This can be inferred by the running of the coupling constants and by the $\U(1)_R$ anomaly charges.

\vskip 0.50cm
\begin{center}
*  *  *
\end{center} 
In the next subsection we review some basic information that shall be needed in the rest of the paper. In Subsection \ref{modulispace}, we introduce the moduli space, and in particular the Higgs branches for the various theories. In Subsection \ref{Coulomb}, we recall the Seiberg-Witten solution for the dynamic  on  Coulomb branch. In Subsection
\ref{formularvacua}, we describe the particular case of the $r$ vacua and the formula for the quark condensate.

\subsection{Moduli Space}
\label{modulispace}

The moduli space of vacua is defined by the $D$-term equations
\bea 
&[\phi,\phi^\dagger] = 0 \ , &  \nonumber \\
& q_a^{\,i} \ q_i^{\,\dagger b} - \tq_a^{\,  \dagger i} \ \tq_i^{\,b} = 0 \ , & \label{Dterm}  
\eea
and the $F$-term equations
\bea
 q_a^{\,i}\  \tq_i^{\, b} &=& 0 \ ,   \nonumber \\
\phi_a^{\,b}\  q_b^{\,i}  &=& 0 \ , \nonumber \\
\tq_i^{\,b}\ \phi_b^{\,a} &=& 0 \ .   
\eea      
From now on, we shall choose all the masses $m^i$ to be equal to zero.

The moduli space can be divided into two different branches. The first is the Coulomb branch where the quarks condensate vanishes  and the gauge group preserves its rank. From the first of (\ref{Dterm}), it follows that the adjoint field $\phi$ can be diagonalized by a gauge transformation. The Coulomb branch has  complex dimension $n_c$ and can be parameterized by the gauge invariant coordinates
\beq
\label{coordinatescoulomb}
u_j=\frac{1}{j} \langle \Trnc \phi^{\,j} \rangle\ , \qquad  j=1,\dots,n_c \ .
\eeq 
The Coulomb branch is a Kahler manifold and develops singularities where some particles, vector multiplets or matter hypermultiplets, become massless. The metric is modified by quantum correction, and the solution is encoded in the Seiberg-Witten curve.

Higgs branches are conic hyper-Kahler manifolds that develop from some of the singularities of the Coulomb branch where two or more flavors of quarks become massless.  Two are the important properties of the Higgs branches: {\it 1)} the metric does not have any dependence on $\phi$;  {\it 2)} the metric does not receive quantum corrections.  Quantum corrections can only change the point on the Coulomb branch where the Higgs branch develops and the pattern of intersections between the various branches; they cannot modify the Higgs branch itself \cite{Argyres:1996eh}.

There is no baryonic branch since we are working in $\U (n_c)$ rather than $\SU (n_c)$.
There are non-baryonic branches and they develop where 
\beq \phi = (0, \dots, 0, \phi_{r+1}, \dots, \phi_{n_c}) \label{rsubmanifold} \ .\eeq After a gauge and flavor rotation, the quarks can be brought in the form
\bea
&&
q = \left(\begin{array}{cccccccc}
k_1 &&& 0\phantom{{}_r} &&& 0      &\\
&\ddots   && &\ddots     && &\ddots \\
&&k_r & &&0\phantom{{}_r} & &       \\
&&&          &&&            &       \end{array}\right)\ , \qquad 
\tq^{\,t} = \left(\begin{array}{cccccccc}
0\phantom{{}_r} &&& k_1 &&& 0      &\\
&\ddots     && &\ddots   && &\ddots \\
&&0\phantom{{}_r} & &&k_r & &       \\
&&&          &&&            &       \end{array}\right) \ , \nonumber \\  
&& k_i \in  \mathbb{R}^+ \ ,
\eea
where $r\leq [n_f/2]$. These non-baryonic $r$ branches are the ones that develop from the sub-manifold (\ref{rsubmanifold}) of the Coulomb branch. The $r$ Higgs branch preserves a $\U(n_f -r)$ subgroup of the flavor symmetry. Part of the flat directions of the Higgs branch are given by the Goldstone bosons of the broken global symmetries. The complex dimension of the branch is $2r(n_f -r)$. Note that a Higgs branch can contain, as a sub-manifold, all the other branches with smaller $r$.


\vskip 0.10cm
For the moduli space of pure $\N=1$ super-QCD, there is only the $D$-term
\beq 
 q_a^{\,i}\  q_i^{\, \dagger b} - \tq_a^{\, \dagger i}\  \tq_i^{\,b} = 0 \ .
\eeq
Up to gauge and global transformation, the quarks can be brought in the form\footnote{See \cite{Intriligator:1995au} for a complete review.}
\bea
&&
q = \left(\begin{array}{ccccccc}
k_1 & & &   & 0\phantom{_1}      &&\\
& k_2    &     && &\ddots &\\
&&   \ddots   &  & &      & 0\phantom{_{n_f -n_c}}\\
&&&     k_{n_c}     &      &      &       \end{array}\right)\ , \qquad 
\tq^{\,t} = \left(\begin{array}{ccccccc}
\tk_1 & & &   & \lambda_1       &&\\
& \tk_2    &     && &\ddots &\\
&&   \ddots   &  & &       & \lambda_{n_f - n_c }\\
&&&     \tk_{n_c}     &            &     &  \end{array}\right) \ , \nonumber \\
&& k_i, \tk_i,  \lambda_i \in \mathbb{R}^+ \ , \nonumber \\
&&|k_i|^2 = |\tk_i|^2 + |\lambda_i|^2\ , \quad \forall i \ .
\eea
The complex dimension is $2n_c n_f - n_c^2$. Note that now there is also a baryonic branch that was absent in the $\N=2$ case.

\vskip 0.10cm
As in the Abelian case, we can break $\N=2$ with a mass term $\mu \Tr \Phi^2/2$ and flow from $\N=2$ to pure $\N=1$.  So to summarize here what happens, {\it this time only classically}, to the moduli space;
\begin{itemize}
\item  The $\N=2$ theory has a Coulomb branch with complex dimension $n_c$ and an equal number of massless vector multiplets. Non-baryonic Higgs branches labeled by $r$ develop from singularities of the Coulomb branch. They are hyper-Kahler manifolds. In particular at the origin $\phi=0$, the maximal Higgs branch $r= [n_f /2]$ develops.  It is given by the quark expectation values $q$ and $\tq$, minus the $F_{\phi}$ conditions, divided by the complexified gauge group $G^{\rm c}$
 \beq
 \mathbb{C}^{2n_c n_f} -  \{ \tq^{\,a}_i \   q^{\, i}_b =0 \} / G^{ \rm c } \ .
 \eeq
 The quaternionic dimension of the maximal Higgs branch is $[n_f/2] n_f -[n_f/2]^2$.
\item  The mass term $\mu \Tr \Phi^2 /2$ in the superpotential lifts all the Coulomb branch with the exception of the origin $\Phi=0$ and the maximal Higgs branch attached to it. 
\item  At $\mu \to \infty$ the $F_{\phi}$ conditions loses its effectiveness. The Higgs branch is thus parameterized by
 \beq
 \mathbb{C}^{2n_c n_f} / G^{\rm c} \ .
 \eeq
  The Higgs branch is enhanced to the $2 n_c n_f -n_c^2$ complex dimensional space.
 We thus smoothly recover the moduli space of pure $\N=1$ SQCD.
\end{itemize}

\subsection{Quantum Dynamics on the Coulomb Branch}
\label{Coulomb}

We now recall some basic properties of the quantum solution of the Coulomb branch given by the Seiberg-Witten curve \cite{Seiberg:1994aj}. 
First we consider $\SU (n_c)$ with $n_f$ flavors.
The curve $\Sigma_{\snd}$  is the Riemann surface
\beq
\label{SigmaN=2}
{y}^2=\frac{1}{4} \det(z-\Phi)^2-\Lambda^{2n_c-n_f}z^{n_f} \ ,
\eeq
double cover of the $z$ plane with genus $g=n_c -1$.  At low energy, one has $n_c-1$ $\U (1)$ gauge multiplets, and we call their scalar components $a_i$, where $i=1,\dots,n_c-1$. The moduli space is a $n_c-1$ dimensional complex manifold ${\cal M}_{\mbox{\tiny $ \SU (n_c)$ }}$, parameterized by the gauge invariant coordinates $u_j$ given in (\ref{coordinatescoulomb}). The SW solution is expressed as the function of the coordinates $s_k$\footnote{ The relationship between $u_j$ and $s_k$, important in what follows, is  encoded in  a single one: $P_{n_c}(z)= z^{n_c} \exp{\left(-\sum_{j=1}^{\infty} \, \frac{u_j}{z^j} \right)_+}$,  
whereby $(\;)_+$ we mean that we discard the negative power expansion.} given by
\beq
 P_{n_c}(z)=\sum_{k=0}^{n_c} \, s_k z^{n_c-k} \ , 
\eeq
\beq
 s_0=1\ , \quad s_1=0 \ , \quad s_k=(-)^k \sum_{i_1 <\dots < i_k} \, \phi_{i_1} \dots \phi_{i_k} \ .
\eeq
Since $\Sigma_{\snd}$ is a genus $n_c-1$ Riemann surface, we can choose $n_c-1$ independents holomorphic differentials:
\beq
\lambda_j \propto \frac{ z^{n_c-j} dz }{y}\ , \qquad j=2,\dots, n_c \ .
\eeq
Each $a_i$ corresponds to an $\alpha_i$ cycle on $\Sigma_{\N=2}$, while its dual $a_{Dj}$ corresponds to a $\beta_j$ cycle chosen in such a way that the intersection is $\langle \alpha_i,\beta_j \rangle =\delta_{ij}$. The SW solution is given by the period integrals
\beq
\label{Solution}
\frac{\de a_i}{\de s_j}=\oint_{\alpha_i}\lambda_j\ ,\qquad
\frac{\de a_{Di}}{\de s_j}=\oint_{\beta_i}\lambda_j\ .
\eeq

Now we study the $\U (n_c)$ theory with $n_f$ flavors, and we will see that the solution can be easily incorporated into the previous ones, with a few modifications. The low-energy theory has one more $\U (1)$ factor that comes  from the decomposition $\U (n_c)= \SU (n_c)\times \U (1)/\Z_{n_c}$, and we denote its scalar component with $a_{n_c}$. This factor has no strong dynamics: in the $n_f=0$ case, it is completely free, while in the $n_f \neq 0$ case, it is infrared free. The moduli space ${\cal M}_{\mbox{\tiny $ \U (n_c)$ }}$ has one dimension more and is parameterized by $u_1$ in (\ref{coordinatescoulomb}).
The Riemann surface is the same given  in (\ref{SigmaN=2}), but here $\phi$ can have non-zero trace and $\Sigma_{\snd}$ also depends  on the modulus $u_1$. To complete our task, we must find the cycle $\alpha_{n_c}$ that corresponds to $a_{n_c}$ and the differential $\lambda_1$ that corresponds to $s_1$. The cycle $\alpha_{n_c}$ is the one that encircles all the cuts in the $z$ plane. Note that this is a trivial cycle, and only a meromorphic differential can be different from zero when it is  integrated around it. The differential that corresponds to  $s_1=-u_1$ is
\beq
\lambda_1 \propto \frac{ z^{n_c-1} dz}{y} \ ,
\eeq
and  is meromorphic because it has a pole at $\infty$.   With these  modifications, the solution is encoded in (\ref{Solution}).

The $\U(n_c)$ theory can also be considered as  part of a bigger, asymptotically free gauge theory. For example, we can take a $\SU (n_c+1)$ $\N=2$ and break it to $\SU (n_c) \times \U(1) / \Z_{n_c}$ at a certain energy scale. This can be achieved simply by choosing  one diagonal element of $\phi$ very distant from the others and fix it. We have thus a sub-manifold of the $\SU(n_c +1)$ moduli space, which is exactly like the one of $\U(n_c)$. What we achieve  with this is that now we can display also the magnetic cycle $\beta_{n_c}$. This is important if we want to compute the monodromies around the singularities and consequently the charges of the massless hypermultiplets on these singularities. We shall use this trick in the next section.

\subsection{Classical and Quantum $r$ Vacua}
\label{formularvacua}

We now review the physics of color-flavor locked vacua or, as we like to call them,  $r$ vacua. The adjoint field $\phi$ has $r_{\cl}$ diagonal elements locked to the hypermultiplet mass $m$ 
\beq
\label{phicondensate}
\langle\phi\rangle=\left(\begin{array}{cccc}
m {\bf 1}_{r_{\cl}}&&&\\
&a_1 {\bf 1}_{n_1}&&\\
&&\ddots&\\
&&&a_k {\bf 1}_{n_k}\\
\end{array}\right)\ ,
\eeq
and the remaining diagonal blocks are divided between the roots of $W^{\prime}$ and $\sum_{j=1}^{k} \, n_j + r_{\cl} = n_c $. We use $r_{\cl}$ to denote the classical value and  $0 \leq r_{\cl} \leq n_{f}$. 
The gauge group is broken by $\langle \phi \rangle$ down to $ \U(r) \times \prod_{j=1}^{k} \U(n_j)$.
If $r$ colors and flavors are locked at the same eigenvalue,  then in the low energy we also have  a massless hypermultiplet in the fundamental of U$(r)$, that we denote as $q$. Apart from these perturbative objects, there are also non-Abelian and Abelian monopoles, each carrying magnetic charge under two of the unbroken gauge groups.

The $F_{\phi}$ term and the $D$ term together yield the potential for the quark fields
\beq
V= g^2 \Tr |q\tq - W^\prime|^2+\frac{g^2}{4}\Tr (qq^\dagger -\tq^\dagger \tq)^2\ .
\eeq
The non-Abelian quarks develop thus a condensate
\beq
\label{classicalcondensate}
\tq q = r_{\cl} W\p(m) \ .
\eeq
The non-Abelian group $\U(r)$ is Higgsed, non-Abelian vortices are formed, and non-Abelian monopoles confined \cite{Pisa}.

Quantum mechanically, is a little bit different. One difference is that in the strong coupling region, where $m$ is of order of the dynamical scale, the massless quarks are continuously transformed into magnetic degrees of freedom, dual-quarks.  Another difference is that (\ref{classicalcondensate}) is modified by quantum corrections. 
We are now going to compute these quantum corrections as in \cite{Bolognesi:2004da}. The quantum parameter $r$ is given by 
\beq
{r}={\rm min}(r_{\cl},n_f-r_{\cl}) \ .
\eeq
\TABLE{
\begin{tabular}{ccccc}
& $\SU(n_f)$ & $\SU({r})$ & U$(1)_0$ & U$(1)_{1,\dots,n_c-{r}}$ \\
\hline
&  $\mathbf{n_f}$&$\mathbf{r}$&$1$&$0$\\
\hline
\end{tabular}
\caption{Low energy in $r$ vacua.}
\label{low}}
There are various reasons for the appearance of this mirror symmetry and the quantum parameter $r$. We saw from the classical discussion of the moduli space that the hyper-Kahler dimension of the $r$ Higgs branch is $r(n_f -r)$. The gauge theory is, classically, $\N=2$ $\SU(r)$ with $n_f$ quarks. Only for $r \leq [n_f /2]$, this is not strong in the infrared, and we can conclude that it remains as a low-energy effective description. Branches which classically have $r_{\cl}$ greater than $[n_f/2]$, quantum mechanically falls into the $r$ vacua classification with $r=n_f - r_{\cl}$.
The factorization of the  $\Sigma_{\snd}$ curve is \cite{Cachazo:2002ry,Cachazo:2003yc}
\beq
\label{factorU(2)}
 \qquad y^2=\frac{1}{4 {g_{n_c - {r}}}^2}(W^{\prime 2}+f)(z-m)^{2 r} \ .
\eeq
It is easy to see, from the curve (\ref{SigmaN=2}), that the factor $z^{2r}$ can be pulled out only if $r \leq [n_f /2]$.  
In the low energy, we have a $\N=2$ $\SU({r}) \times \U(1) \times \U(1)^{n_c-{r}}$ gauge theory with hypermultiplet $\tD, D$ with charges given in Table \ref{low}. 
The low-energy superpotential is
\beq
\W_{\rm low}=\sqrt{2}\left( \sum_{i=1}^{n_f}\, ( \tD_i A_{\mathbf{r}} D^i +  \tD_i A_{0} D^i) - W_{\mathrm{eff}}(A_{\mathbf{r}},A_0,\dots,A_{n_c-r}) \right)\ ,
\eeq
where the effective superpotential that breaks to $\N=1$ is
\beq
W_{\mathrm{eff}}=\sum_{j=0}^{n_c - r} \, g_j u_{j+1}(A_{\mathbf{r}},A_0,\dots,A_{n_c-r}) \ .
\eeq
We choose a superpotential with $k=n_c-r$, so that all the possible points in the $r$ sub-manifold can in principle be selected. $A$'s are the chiral superfields of the $\N=2$ gauge multiplets, $A_{\mathbf{r}}$ is  one of the $\SU({r})$ gauge multiplet, and $A_0$ is one of the U$(1)$ multiplet coupled to the dual quarks.

To compute the quantum condensate $\td d$, it is more convenient to split  the flavors' masses a little bit and use the Abelianized theory so obtained. Then we simply make the limit of coincident masses for the result so obtained. 
We have $r$ flavors with masses $m_j$ with $j=1,\dots,r$. In the limit if coincident masses $m_j \to m$, $\forall j$ and we recover the $r$ vacuum. The chiral superfields that we previously called $A_{\mathbf{r}}$ and $A_0$, are now described by $r$ Abelian fields that we denote as $A_{(0,j)}$ with $j=1,\dots,r$. Each $\U(1)_{(0,j)}$ is locked to one flavor $\tD_j$, $D_j$.
 The low-energy superpotential is now  
\beq
\label{lowsup}
\W_{\rm low}=\sqrt{2}\left(\sum_{j=1}^{r}\, \tD_j A_{(0,j)} D^j - W_{\rm eff}(A_{(0,1)},\dots,A_{(0,r)}; A_{1}, \dots, A_{n_c-r}) \right) \ .
\eeq
For our computation,  we need to consider only the $F_A$ terms of the potential:
\bea
F_{A_{(0,j)}}=2 e_{(0,j)}^2  \left|\td_j d^j - \frac{\de W_{\rm eff}}{\de a_{(0,j)}} \right|^2\ , &\qquad&
j=1,\dots,r \ , \nonumber \\
F_{A_s}=2 e_s^2 \left|\frac{\de W_{\rm eff}}{\de a_s}\right|^2\ , &\qquad& s=1,\dots,n_c-r \ . 
\eea
The first ones give the condensates, while the second gives the stationary condition necessary to compute the position in the moduli space:
\bea
\td_j d^j = \frac{\de W_{\rm eff}}{\de a_{(0,j)}} \ , &\qquad&
j=1,\dots,r \ , \nonumber \\
\label{formulas} 
0 = \frac{W_{\rm eff}}{\de a_s}\ , &\qquad& s=1,\dots,n_c-r \ . 
\eea
We can write a matrix equation like (\ref{formulas}) where the couplings vector is
\beq
\left[{\bf g}\right]=(g_0, \dots, g_{n_c-r}, 0, \dots, 0) 
\eeq
and the tension vector is 
\beq
\left[{\bf \td d}\right]=(\td_1 d_1,\dots, \td_r d_r,  0 , \dots,  0) \ .
\eeq
With these conventions, the equation (\ref{formulas}) becomes
\beq
\label{generalmatrice}
\left[{\bf \td d}\right] = \left[\mathbf{\frac{\de u}{\de  a}  }\right]  \left[{\bf g}\right] \qquad \Longrightarrow \qquad
\left[{\bf g}\right]= \left[\mathbf{\frac{\de a}{\de  u}  }\right] 
\left[{\bf \td d}\right] \ .
\eeq
where in the last we have simply multiplied by the inverse matrix. The last version is the one that we shall find more convenient for computing the condensates.

Let us begin with the simplest example: the $r=1$ vacuum of the $(n_c,n_f) = (2,2)$ theory. 
The superpotential is quadratic
\beq
\label{superoptentialexample}
W(z)=g_0z+\frac{g_1}{2}z^2\ ,
\eeq
and the factorization of $\Sigma_{\snd}$ gives
\bea
y^2 &=& \frac{1}{4} P_2 (z)^2 -\Lambda^2 (z-m)^2 \nonumber \\  
&=& \frac{1}{4 g_1^2}(W^{\prime 2}+f)(z-m)^2\ .
\eea
The cycle $\alpha_0$ is the one that encircles the double roots $z=m$. The crucial ingredient for the following computation is the residue  around the cycle $\a_0$: 
\beq
\label{residue1}
\frac{1}{4 \pi i} \oint_{\a_0} \frac{dz}{y}=\frac{g_1}{\sqrt{W^{\prime 2}+f}} \ .
\eeq
In this case, the relationship between $s$ and $u$ coordinates is:
\beq
\label{surelation}
s_1=-u_1 \ , \qquad  s_2 = - u_2 +\frac{{u_1}^2}{2} \ .
\eeq 
For our proof,  we will need  to calculate only
\beq
\frac{\de a_0}{\de u_2}=\frac{\de a_0}{\de s_1}\frac{\de s_1}{\de u_2}+\frac{\de a_0}{\de s_2}\frac{\de s_2}{\de u_2}=-\frac{\de a_0}{\de s_2} .
\eeq
First, we observe that the solution (\ref{Solution}) and the residue (\ref{residue1}) give\footnote{The proper normalization of the holomorphic differential has been chosen to reproduce the correct semiclassical result}
\beq
\label{op}
\frac{\de a_0}{\de s_2}=-\frac{g_1}{\sqrt{W^{\prime 2}+f}} \ .
\eeq
Then, writing the equations (\ref{formulas}) in the matrix form suggested in (\ref{generalmatrice}), we get
\beq
\label{matrice}
\left(
\begin{array}{c}
g_0\\
g_1\\
\end{array}
\right)
=
\left(
\begin{array}{cc}
\,\de a_0/\de u_1\,&\,\de a_1/\de u_1\,\\
\,\de a_0/\de u_2\,&\,\de a_1/\de u_2\,\\
\end{array}
\right)
\left(
\begin{array}{c}
\td d\\
0\\
\end{array}
\right)\ .
\eeq
The simple passage of multiplying by the inverse matrix has simplified  our work a lot because now (\ref{matrice}) is expressed as a function of  $\de a_i/\de u_j$, known through (\ref{Solution}).  Furthermore, only $\de a_0 / \de u_{1,2}$, the ones obtained by an integral around the collided roots,  are important because the others are multiplied by zero.
From (\ref{matrice}), we need only the second equation
\beq
\label{lastone}
g_1=  \frac{\de a_0}{ \de u_2}  \td d  \ ,
\eeq
that, using (\ref{op}), gives the condensate
\beq
\label{strongtension}
\td d=\left.\sqrt{W^{\prime 2}+f}\right|_{z=m}\ .
\eeq

We give another example $\U(3)$ with two flavors of mass $m_1$ and two flavors of mass $m_2$ (see Figure \ref{cycles} for the roots and cycles). 
\FIGURE{
\includegraphics[width=24em]{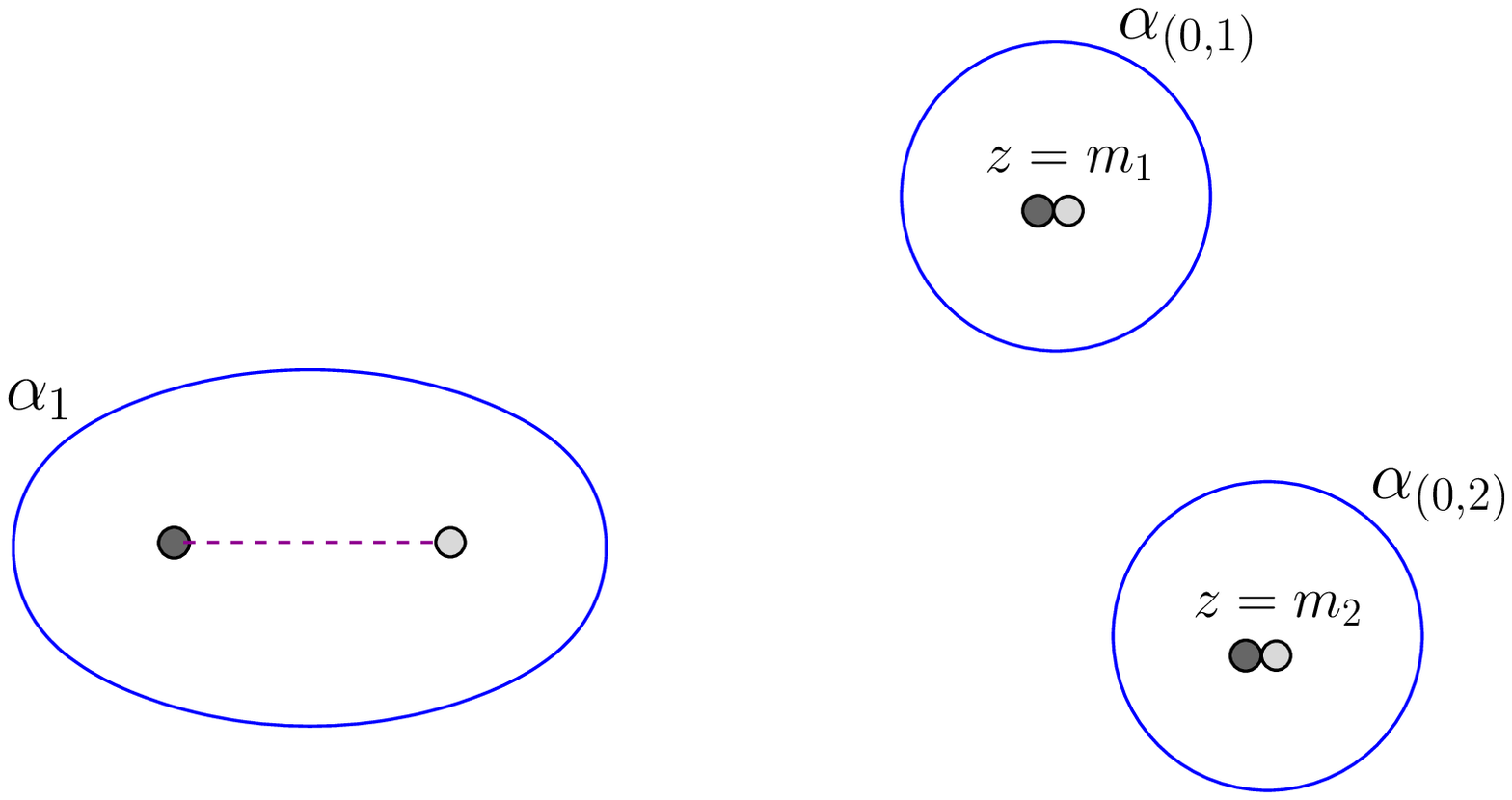}
\caption{Cycles in $\U (3) \to \U(1)_{(0,1)} \times \U (1)_{(0,2)} \times \U (1)_1$ theory. Displayed are the electric cycles $\alpha_{(0,1)}$, $\alpha_{(0,2)}$, and $\alpha_1$. Two flavors at the poles $m_1$ and $m_2$ are massless and charged with respect to $\U(1)_{(0,1)}$ and $\U(1)_{(0,2)}$. In the limit $m_1 \to m_2 \to m$, we recover the $r=2$ vacuum by breaking $\U(3) \to \SU(2) \times \U(1)_0 \times \U(1)_1$.}
\label{cycles}
}
The superpotential is still (\ref{superoptentialexample}),
while the factorization gives $\Sigma_{\snd}$:
\bea
y^2 &=& \frac{1}{4} P_3 (z)^2 -\Lambda^2 (z-m_1)^2 (z-m_2)^2 \nonumber \\  
&=& \frac{1}{4 g_1^2}(W^{\prime 2}+f)(z-m_1)^2(z-m_2)^2\ .
\eea
The relationship between $s$ and $u$ in this case is given by (\ref{surelation}) plus:
\beq
s_3=-u_3+u_1 u_2 - \frac{{u_1}^3}{6} \ .  
\eeq 
By using the trick explained before, we write the last of (\ref{generalmatrice}) for this particular case
\beq
\label{matricedue}
\left(
\begin{array}{c}
g_0\\
g_1\\
0\\
\end{array}
\right)
=
\left(
\begin{array}{ccc}
\, \de a_{(0,1)}/\de u_1 \, &\,  \de a_{(0,2)}/\de u_1 \, & \, \de a_1/\de u_1 \, \\
\, \de a_{(0,1)}/\de u_2 \, &\,  \de a_{(0,2)}/\de u_2 \, & \, \de a_1/\de u_2 \, \\
\, \de a_{(0,1)}/\de u_3 \, &\,  \de a_{(0,2)}/\de u_3 \, & \, \de a_1/\de u_3 \, \\
\end{array}
\right)
\left(
\begin{array}{c}
\td_1 d^1\\
\td_2 d^2\\
0\\
\end{array}
\right)\ .
\eeq
Now, as in the previous case,  we need to  calculate only the  residues around $\tm_1$ and $\tm_2$.
The last equation of (\ref{matricedue}) is
\beq
\label{sistem1}
0 =   \td_1 d^1\frac{1}{\left.\sqrt{W^{\prime 2}+f}\right|_{m_1} (m_1-m_2)   } +   \td_2 d^2\frac{1}{\left.\sqrt{W^{\prime 2}+f}\right|_{m_2}(m_2-m_1)} \ .
\eeq 
 using also the second equation of (\ref{matricedue}). Using
\beq
\frac{\de a}{\de u_2}= -\frac{\de a}{\de s_2} + u_1 \frac{\de a}{\de s_3} \ ,
\eeq
the second equation of (\ref{matricedue}) leads to another independent equation
\beq
1=    \td_1 d^1\frac{ m_1}{\left.\sqrt{W^{\prime 2}+f}\right|_{m_1} (m_1-m_2)   } + \td_2 d^2 \frac{ \tm_2}{\left.\sqrt{W^{\prime 2}+f}\right|_{m_2}(m_2-m_1)}  \ .
\eeq
This equation together with (\ref{sistem1}) are enough to establish the solution
\beq
\label{solutionone}
\td_1 d^1=\left.\sqrt{W^{\prime 2}+f}\right|_{m_1} \ ,\qquad \td_2 d^2=\left.\sqrt{W^{\prime 2}+f}\right|_{m_2}\ .
\eeq

Thus we obtain, in the limit of coincident masses $m_1=m_2=m$, the little formula for the dual-quark condensate
\bea
\label{dualcondensate}
\td d= r \left. \sqrt{ W^{\prime}(z)^2 + f(z)} \right|_{z=m} \ .
\eea
This is the quantum generalization of (\ref{classicalcondensate}). The crucial ingredient for the computation has been  the residue around the mass poles. We have for simplicity restricted ourselves to the specific case of $n=k=n_c -r$, but the formula (\ref{dualcondensate}) still holds in the generic case. We refer to \cite{Bolognesi:2004da} for more details about the derivation.  In Section \ref{MQCD}, we shall re-derive it a simpler way in the MQCD setup.


\section{The First Examples of Quantum Coincidence}
\label{Example}

We now describe the first example of coincidence vacua. We begin with the easiest case, $n_c = 2$.
After diagonalization, we have $\phi={\rm diag} (\phi_1, \phi_2 )$. So the gauge invariant coordinates of the moduli space are $u_1 = \Tr \phi = \phi_1 +\phi_2$ and $u_2 = \frac12 \Tr \phi^2 = \frac12 (\phi_1^2 + \phi_2^2)$.
The SW curve is:\footnote{When we write $\Lambda$ without any subscript, we always mean $\Lambda_{\snd}$.}
\bea
{y}^2&=&\frac{1}{4} (z-\phi_1)^2(z-\phi_2)^2-\Lambda^{4-n_f}z^{n_f} \nonumber \\
&=& \frac{1}{4} \left(z^2 -u_1 z + \frac{u_1^2}{2}-u_2\right)^2-\Lambda^{4-n_f}z^{n_f} \nonumber \\
&=& \left( \frac{1}{2} (z-\phi_1)(z-\phi_2)-\Lambda^{2- n_f /2}z^{ n_f / 2}\right) \left( \frac{1}{2} (z-\phi_1)(z-\phi_2) + \Lambda^{2-n_f / 2}z^{n_f /2}\right) \ . \label{duesw}
\eea
The last passage can be done only for $n_f$ even.

We start now with the specific case of $n_f =2$.  Thanks to the factorization (\ref{duesw}), we can now can explicitly compute the four roots of the polynomial.  We call them $z^-_{\, 1,2}$ and $z^{+}_{\, 1,2}$, respectively, for the left and right factors. Their value is
\bea
z^-_{\, 1,2} &=&  \frac12 \left( u_1+ 2 \Lambda \pm \sqrt{( u_1+ 2 \Lambda)^2-4\left( \frac{u_1^2}{2} - u_2 \right)} \right) \ ,  \nonumber \\ z^+_{\,1,2}&=& \frac12 \left( u_1- 2 \Lambda \pm \sqrt{( u_1-2 \Lambda)^2-4 \left( \frac{u_1^2}{2} - u_2 \right)} \right) \ .
\eea

It is good to begin to  study  the problem in different sections of the moduli space.  First, at $\Tr \phi =0$, where it corresponds to the $\SU(2)$ moduli space. The curve is 
\bea
{y}^2&=& \frac{1}{4} (z^2  -u_2)^2-  \Lambda^2 z^{2} \nonumber \\
&=& \left( \frac{1}{2} ( z^2 -u_2 )-\Lambda z \right) \left(  \frac{1}{2} (z^2 -u_2) + \Lambda z \right) \ .
\eea
There are two singular points. One is $u_2 =0$ where $z^-_{\,1,2} = 2\Lambda, 0  $ and $z^+_{\,1,2}=0, -2\Lambda $ (Figure \ref{nbbroots}). This is the root of the $r=1$ branch.  The other singularity is at $u_2 = - \Lambda^2$ and  $z^-_{\,1,2} = \Lambda ,\Lambda   $ and $z^+_{\,1,2}=-\Lambda,-\Lambda $ (Figure \ref{bbroots}). This is the root of the $r=0$ branch (this is what is called ``root of the baryonic branch'' in \cite{Argyres:1996eh}). Note that the singularity is doubled, and looking at the complete moduli space, it will be clear why.
\DOUBLEFIGURE{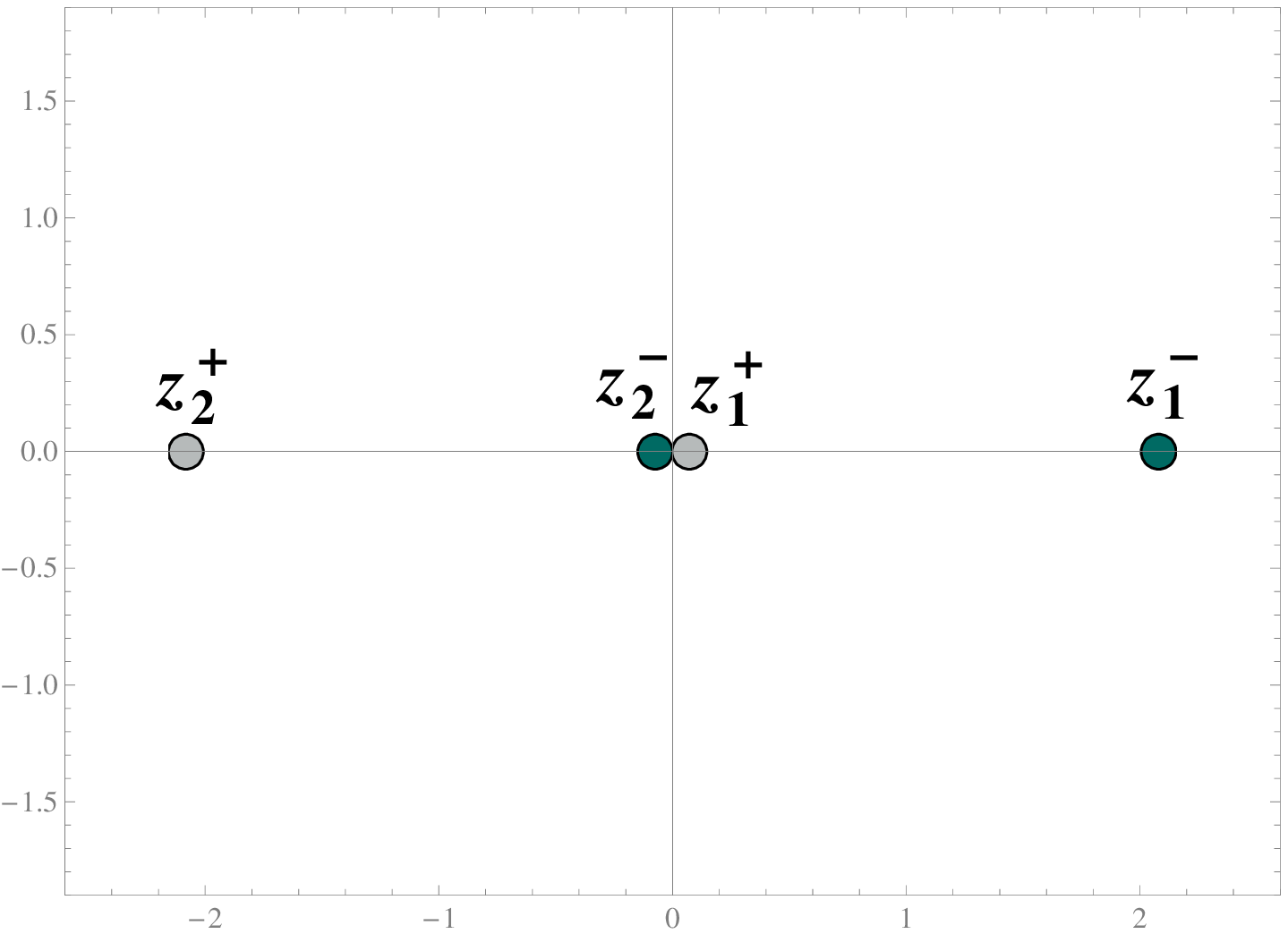,width=17.5em }{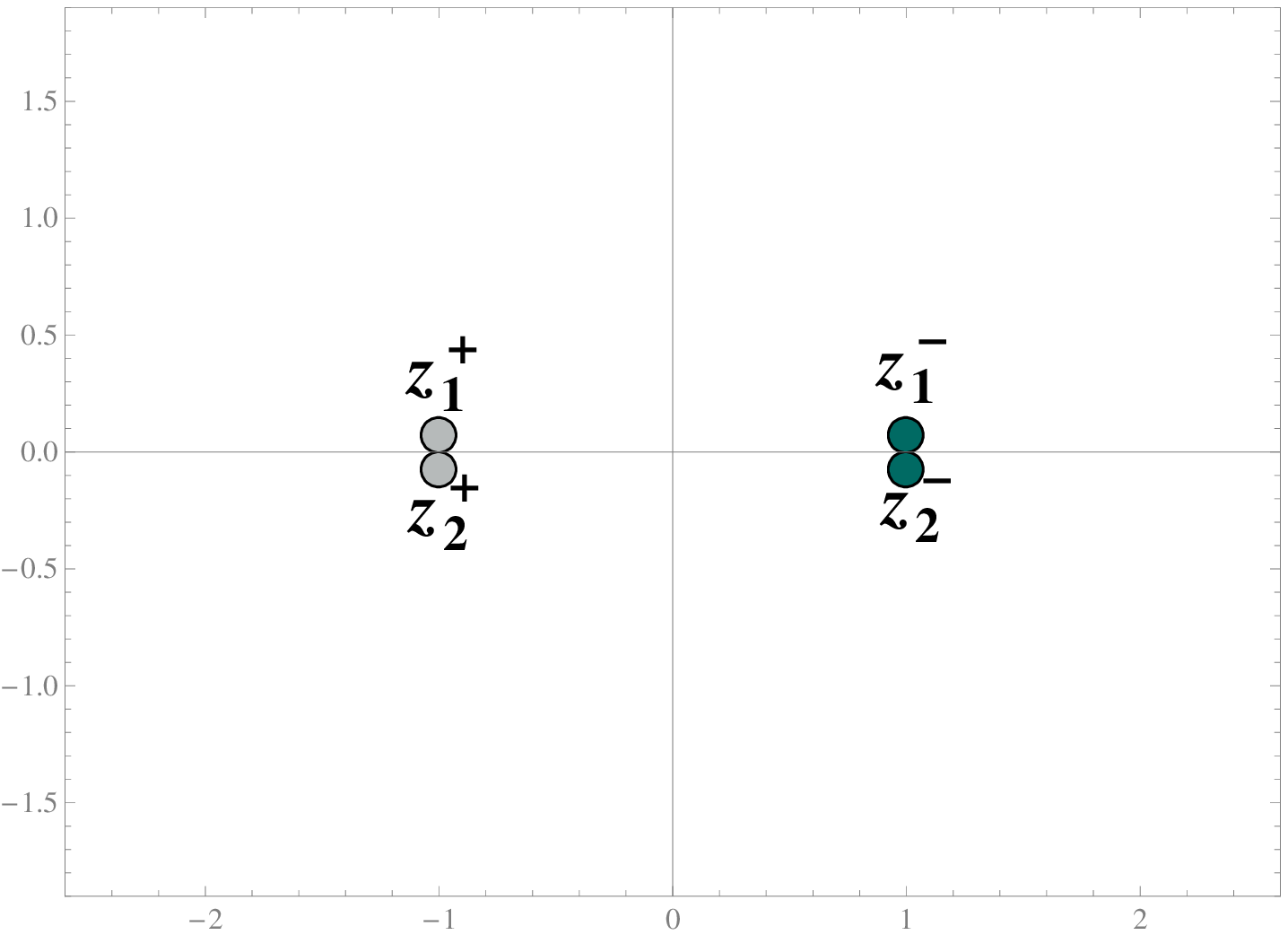,width=17.5em }{Roots near the $r=1$ vacuum: $u_1=0$, $u_2=.16$.\label{nbbroots}}{Roots near the $r=0$ vacuum: $u_1 = 0$, $u_2=-1.006 $.\label{bbroots}}

Now we take a section where $\Tr \phi = M \gg \Lambda$.   One singularity is at $\phi={\rm diag} (0, M)$,  where one quark becomes massless. It corresponds to $u_2= M^2$. The curve is 
\bea
{y}^2&=& \frac{1}{4} \left(\det \mbox{\small $\left(\begin{array}{cc}0 &  \\ 
 & M 
\end{array}\right)$ }  \right)^2-\Lambda^2 z^2 \nonumber \\
&=& z^2\left( \frac{1}{4}(z-M)^2-\Lambda^2\right)  \ .
\eea
The other singularities are ``near''  $\phi={\rm diag}(M/2,M/2) $
and are strong coupling singularities of a pure $\SU(2)$ gauge theory. They correspond to, respectively, to the coincidence of $z^-_{1} = z^-_{2}$ at $u_2=-\Lambda^2 /4 - M\Lambda/2$, and the coincidence of  $z^+_{1} = z^+_{2}$ at $u_2=-\Lambda^2/4 + M\Lambda/2$. They are the usual monopole and dyon singularity of pure $\SU(2)$.

\FIGURE{
\includegraphics[width=22em]{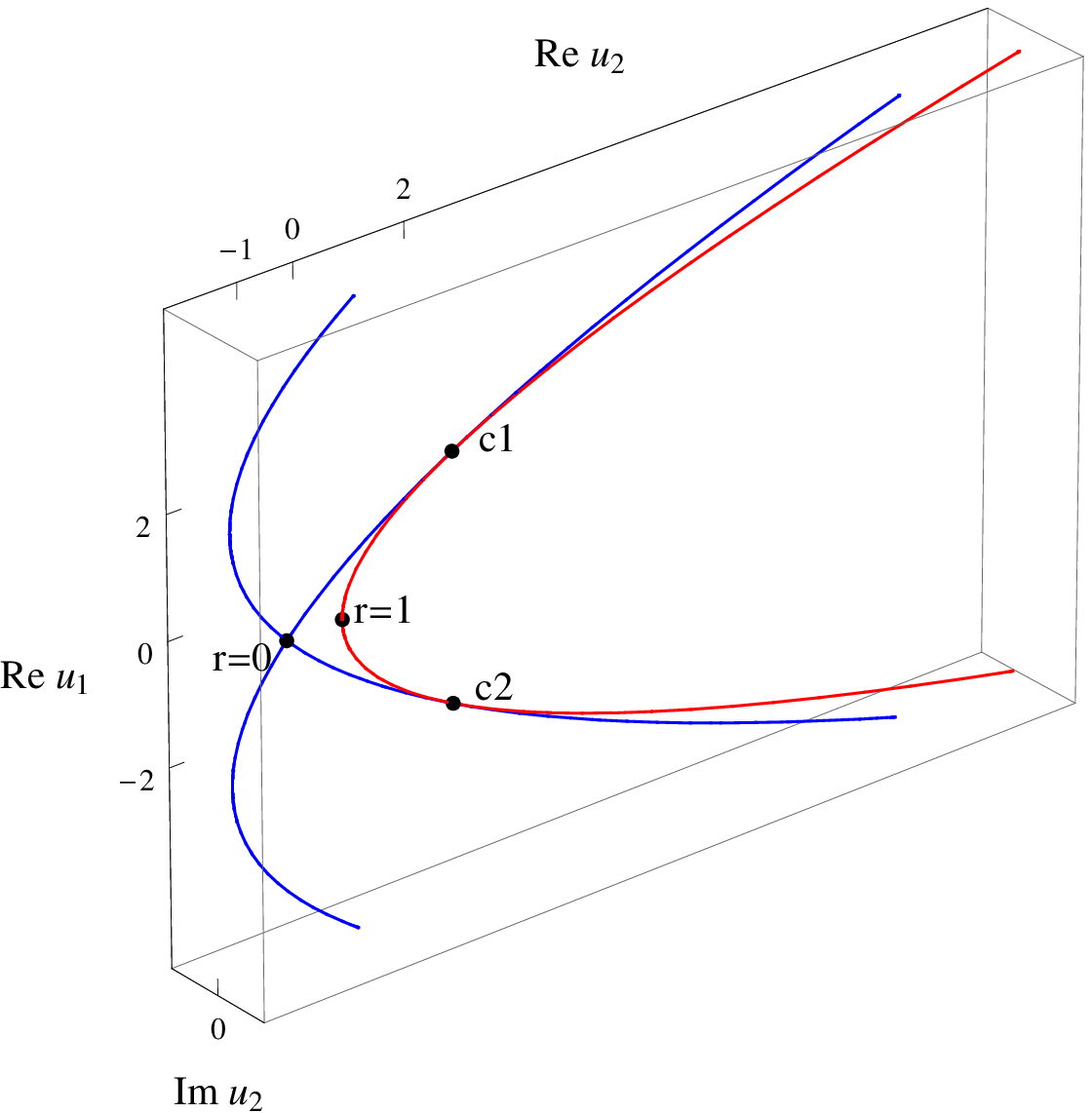}
\caption{The $\Im u_1 =0$ section of the moduli space. The lines are the co-dimension two singularities. The four points are the $r=0,1$ vacua at $u_1=0$, and the two coincidence points.}
\label{plot}
}
We can now draw a picture with the singularities (Figure \ref{plot}). The space $u_1,u_2$ is two complex dimensional. Singularities are one complex dimensional surfaces, objects of co-dimension one, where two roots of the SW curve coincide.  We draw only three real dimensions, the plane $u_2$ and the real part of $u_1$. In this plot, the singularities are lines (real co-dimension two).
The three singular curves are
\bea
 u_2 &=& - \Lambda^2 \pm u_1 \Lambda + \frac{u_1^2}{4} \ , \label{bbsingularity} \\
 u_2 &=& \frac{u_1^2}{2} \ .\label{nbbsingularity}
\eea
There are three special points where the curves intersect, and there are more coincidences of roots. One has already been discussed and is the root of $r=0$ branch at $u_1 =0$. 
The other two special points intersections are at $u_1 = \pm 2 \Lambda$ and $u_2= 2 \Lambda^2 $. Where the curve is
\beq
y^2 = z^3 (z \pm 4 \Lambda) \ .
\eeq
\DOUBLEFIGURE{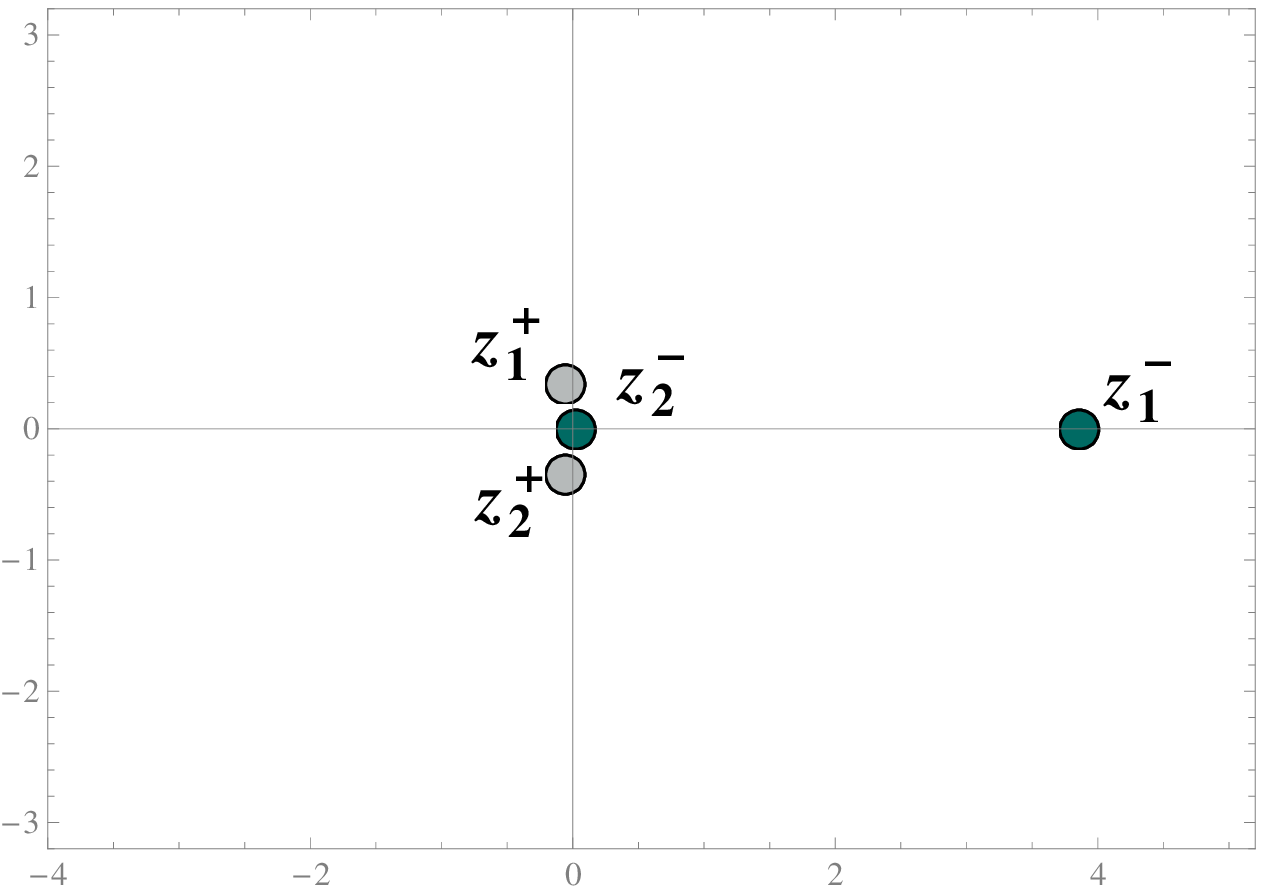,width=17.0em}{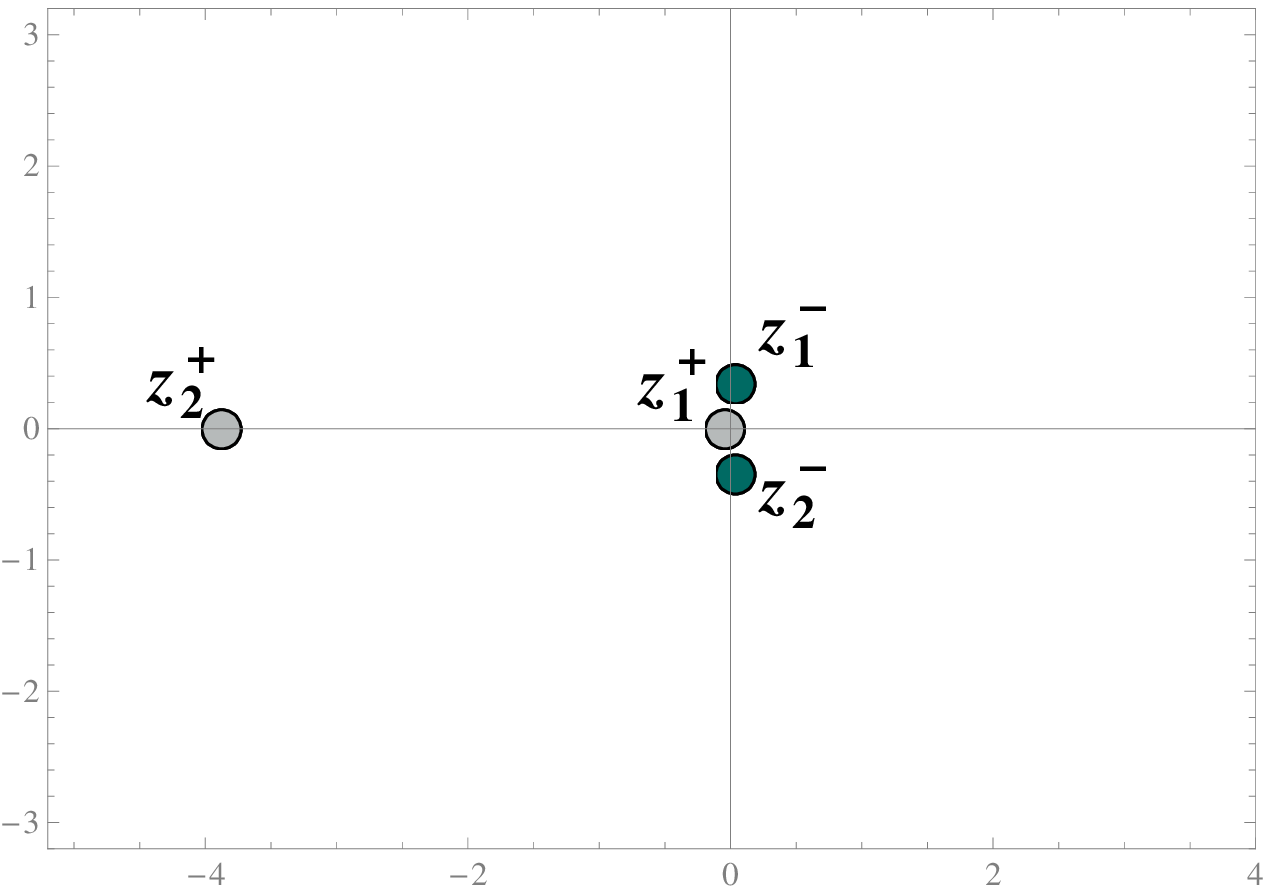,width=17.0em}{Roots near the first coincidence vacuum: $u_1= 1.9$, $u_2= 1.68$. \label{c1}}{Roots near the second coincidence vacuum: $u_1= -1.9$, $u_2= 1.68$. \label{c2}}

Clearly something special happens in these two singularities (see Figures \ref{c1} and \ref{c2} for the roots in these vacua). These are the coincidence points. To understand what is happening,  it is good to start from the $r=1$ vacuum at $u_1 =0$ of Figure \ref{nbbroots}. Here we have two roots in zero, and they correspond to a massless quark. The other two roots are in $\pm 2 \Lambda$. A mass perturbation $W=\mu \Tr \Phi^2/2$ selects this $r=1$ vacuum, and the two unpaired roots correspond to the polynomial $W^\prime(z)^2 + f(z)$. The splitting is due to the quantum effect of $f(z)$.  
We can now change the $u_1$ coordinate, while remaining on the singularity $u_2=u_1^2 /2$, and so change the position of the unpaired roots. To achieve this, we can use a generic quadratic superpotential
\beq
\label{superpotentialgeneric}
W(\Phi) = \mu \Tr \left(\frac{\Phi^2}{2} - \alpha \Phi \right) \ .
\eeq
The parameter $\alpha$ determines the level of $u_1$ that this superpotential selects. It selects all the maximal singularity points at the level $u_1=\alpha$.\footnote{Consider the quark  singularity $u_2=u_1^2 / 2$. If we want to select a generic point on this singularity then $W = \mu (u_2 - \alpha u_1)= \mu ( u_1^2/2 - \alpha u_1)$. The vacuum selected will be thus at $u_1 =\alpha$ and $u_2 =\alpha^2/ 2$.} Changing $\alpha$, we can thus move along $u_1$ while staying on the singular lines. In this way, the two unpaired roots change their position while the two at the quark singularity remain fixed. When we reach $u_1 = 2 \Lambda$, the root $z^+_{\, 2}$ collides with the quark roots, and thus we have the coincidence point, exactly as was anticipated in Figure \ref{rcollision}. At $u_1=-2 \Lambda$ is $z^-_{\,1}$ that coincides with the quark roots, at $u_1=+2 \Lambda$ is $z^+_{\,2}$ that coincides with the quark roots.
So we have that the two superpotentials
\beq
\label{superpotentialcoincidence}
W(\Phi) = \mu \Tr \left(\frac{\Phi^2}{2} \mp 2 \Lambda \Phi \right)
\eeq
select respectively the two coincidence vacua.

Now we discuss the low-energy effective action in these various vacua. 
First the two $u_1 =0$ cases.  The root of the $r=1$ non-baryonic Higgs branch  (see  Table \ref{nbbduecharges}) has a flavor doublet $\tD , D$ (the dual-quark) charged under only one of the $\U(1)$'s factors.
The superpotential is
\beq
{\cal W}_{r=1} = \sqrt{2} \tD A_1 D \ ,
\eeq
The Higgs branch emanating from it is exactly the non-baryonic branch $r=1$.  
\bea
\td_i d^i= 0 \ , \nonumber  \\
|\td_i|^2 - |d^i|^2=0  \ .
\eea
This correctly reproduces the $r=1$ non-baryonic Higgs branch.
The low-energy effective action, after the mass perturbation $W=\mu \Tr \Phi^2 / 2$, becomes
\beq
{\cal W}_{r=1}^\prime  = \sqrt{2} \left( \tD A_1 D  - \mu u_2(A_1,A_2) \right) \ .
\eeq
The dual quark develops a condensate, the $\U(1)_1$ is Higgsed, and the only thing that remains in the infrared is the free $\U(1)_2$ theory.
\DOUBLETABLE[h]{\begin{tabular}{c|cccc}
&& $\U(1)_1$ & $\times$ & $\U(1)_2$ \\
\hline
${\bf 2} \times D$ &  & $1$  && $0$  \\
 \hline
\end{tabular}
}{\begin{tabular}{c|cccc}
&& $\U(1)_1$ & $\times$ & $\U(1)_2$ \\
\hline
$E_1$ &  & $1$  && $0$      \\
$E_2$ &  & $0$  && $1$ 
 \\ \hline
\end{tabular}
}{Low energy in the $r=1$ vacuum. \label{nbbduecharges}}{Low energy in the $r=0$ vacuum. \label{bbduecharges}}

Then we have the $r=0$ vacuum whose  particles and charges are given in Table \ref{bbduecharges}.
The low-energy effective action is
\beq
{\cal W}_{r=0}  = \sqrt{2} \left(  \widetilde{E}_1 A_1  E_1  +  \widetilde{E}_2  A_2   E_2  \right) \ .
 \eeq
 There is no Higgs branch emanating from it. 
This point is again selected by $W=\mu \Tr \Phi^2 / 2$, and the low-energy effective action becomes
 \beq
{\cal W}_{r=0}^\prime =\sqrt{2} \left( \widetilde{E}_1 A_1  E_1 + \widetilde{E}_2  A_2  E_2  - \mu u_2(A_0,A_1) \right) \ .
 \eeq
Both hypermultiplets condense, and thus we have a mass gap.


\vskip 0.10cm
We finally come to the  coincidence points.  In these vacua, mutual non-local particles become massless at the same time.  We have to determine exactly the charges of these particles with respect to a common basis (note that the previous two Tables \ref{nbbduecharges} and \ref{bbduecharges} were not in the same basis).   First of all, we choose a basis of cycles as in Figure \ref{cyclesadpoint}. Note that we displayed the roots $z_{\, 1,2}^-$, $z^+_{\, 1,2}$ near to the first coincidence point. There are other two roots in the figure; we need to embed the theory in $\SU(3) \to \SU(2) \times \U(1)$ in order to show the magnetic cycle  $\beta_2$. Cycle $\alpha_1$ winds around the roots $z^-_2$ and $z^+_1$. Its dual $\beta_1$ winds around the roots $z^-_2$ and $z^+_2$. Cycle $\alpha_2$ winds around all the four roots, and its dual $\beta_2$ winds around $z^-_1$, and passes through the two extra roots.
\FIGURE{
\includegraphics[width=22em]{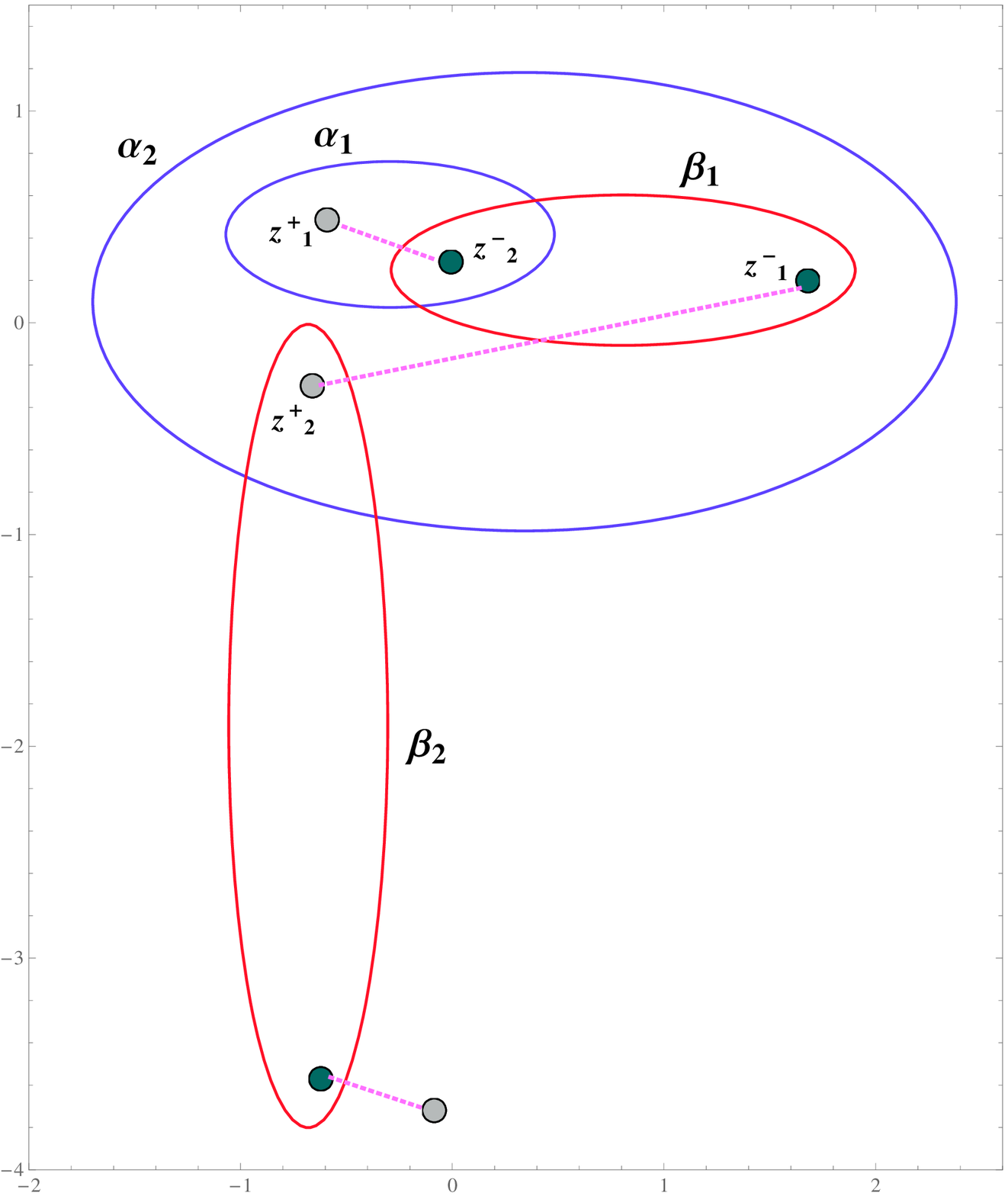}
\caption{Our choice of cycles $\alpha_1, \alpha_2$ and $\beta_1,\beta_2$ near the first coincidence point (Figure \ref{c1}).}
\label{cyclesadpoint}
}

\TABLE[h]{
\begin{tabular}{c|cccc}
&& $\U(1)_1$ & $\times$ & $\U(1)_2$ \\
\hline
${\bf 2} \times  D$ &  & $\phantom{-}1_{\rm ele}$  && $0$  \\
$  B$ & &   $-1_{\rm mag}$ && $1$ \\ \hline
\end{tabular}
\caption{Particles and charges in the coincidence point.}
\label{chargessuperconformal}
}
This is the choice of basis more convenient for the $r=1$ vacuum. A monodromy around the (\ref{nbbsingularity}) singularity gives the charge of the dual quark $\tD$, $D$: a doublet of the flavor group and electrically charged under the $\U(1)_1$.
To get the charge of the other massless particle, we compute the monodromy around the singularity (\ref{bbsingularity}). Choosing the vector of cycles like  $(\alpha_1,\alpha_2|\beta_1,\beta_2)$, the monodromy is
\beq
M=\left(\begin{array}{c|c}  \mathbf{1} -h \otimes q & \phantom{\mathbf{1}}-h  \otimes h \\ \hline \phantom{\mathbf{1} \,} q \otimes q  & \mathbf{1} + q \otimes  h \end{array} \right)  =\left(\begin{array}{cc|cc}   \phantom{-}1&\phantom{-}1&-1&\\&\phantom{-}1&&\\ \hline &&\phantom{-}1&\\&\phantom{-}1&-1&\phantom{-}1 \end{array} \right)
\eeq
This particle, which we call $B$, is a singlet with respect to the color, magnetically charged with respect to $\U(1)_1$, and electrically charged with respect to $\U(1)_2$ (see Table \ref{chargessuperconformal}). This is nothing but one of the two particles of Table \ref{bbduecharges} that are massless in the $r=0$ vacuum. But due to the change of basis, it is now a magnetic object.

This is a particular kind of Argyres-Douglas singularity \cite{Argyres:1995jj,Argyres:1995xn,Eguchi:1996vu}.  The low-energy dynamics consists of a non-local, strongly interacting superconformal field theory.   The superpotentials (\ref{superpotentialcoincidence}) select, respectively, these two coincidence points breaking $\N=2$ down to $\N=1$. No condensate is developed, and this means that the theory in the infrared still remains superconformal.\footnote{The fact that the condensate vanishes in this particular point has also been previously noted in \cite{Gorsky:2000ej}.}  Our claim is that in these points there is an interpolation between $\N=2$ super-QCD and pure $\N=1$ super-QCD.  We shall come back at the end of the next section, after the generalization to arbitrary $n_c,n_f$, for the discussion and interpretation of this important issue. For the moment, let us keep in mind the lessons that we learn from this example: 
\begin{itemize}
\item We have $2n_c -n_f$ ($2$ in this case) coincidence points symmetric under the  $\Z_{2n_c-n_f}$ remnant of the $\U(1)_R$ symmetry.
\item The coincidence vacua lie at the intersections between the $r$ vacua singularities. 
\item Mutual non-local particles become massless in these vacua. They are particular cases of Argyres-Douglas singularities. 
\end{itemize}
Let us now consider a few more specific examples.

\vskip 0.50cm
\begin{center}
*  *  *
\end{center}
Another example is still $n_c=2$ but now with $n_f =3$.  This belongs to a special class $n_f = 2n_c -1$. 
The number $2n_c -n_f$ is particularly important due to the discrete $R$ symmetry. The number of coincidence points is in fact equal to $2n_c -n_f$ since they spontaneously break this discrete symmetry. The previous example had two coincidence points. This example has only one coincidence point, and it lies in the $u_1 =0$ section of the moduli space. The SW curve is
\beq
{y}^2 = \frac{1}{4} \left(z^2 -u_1 z + \frac{u_1^2}{2}-u_2 \right)^2-\Lambda z^3  \ . 
\eeq
Now we cannot use the last passage of (\ref{duesw}) and there is no easy expression for the roots. But the maximal singularity is nevertheless easy to detect. 
The singularity is only one and is in the center of the moduli space at $u_1 =0$, $u_2 =0$. The curve factorizes as
\beq
{y}^2 = \frac{1}{4} z^3(z -4 \Lambda)   \ .
\eeq
So in this case the classical notion of the origin of the moduli space persists. It can be seen that of three singularities of co-dimension two meet at this point. The superpotential that leaves this point is simply $W(\Phi) = \mu \Tr \Phi^2 /2$.

\vskip 0.10cm
Another fully computable example, with different $2n_c -n_f$ from the previous ones, is  $n_c =4$, and $n_f =4$. Now $2n_c -n_f=4$ and we expect four coincidence points.
The non-baryonic branch with $r=2$ has the following curve
\bea
{y}^2 &=&  z^4 \left( \frac{1}{4} \left(z^2 -u_1 z + \frac{u_1^2}{2}-u_2 \right)^2-\Lambda^4 \right) \nonumber  \\
&=& \frac{1}{4} z^4 \left( z^2 -u_1 z + \frac{u_1^2}{2}-u_2 -2 \Lambda^2 \right)\left(z^2 -u_1 z + \frac{u_1^2}{2}-u_2 + 2\Lambda^2 \right) \ .
\eea
The four extra-roots are 
\bea
&&  \frac{u_1 \pm \sqrt{-u_1^2 + 4u_2 + 8\Lambda^2}}{2} \ , \nonumber \\
&&   \frac{ u_1 \pm \sqrt{-u_1^2 + 4u_2 - 8\Lambda^2}}{2} \ .
\eea
Two must collide, and one of the other two, the split ones,  must be at the mass value zero. One choice is
\bea
&& -u_1^2 + 4u_2 - 8\Lambda^2 = 0 \ , \nonumber \\
&&   u_1 = \pm \sqrt{-u_1^2 + 4u_2 + 8\Lambda^2} \ ,
\eea
whose solution is $u_1=\pm 4\Lambda$ and $u_2= 6\Lambda^2$.
The other is
\bea
&&  -u_1^2 + 4u_2 + 8 \Lambda^2 =0 \ , \nonumber \\
&&   u_1= \pm \sqrt{-u_1^2 + 4u_2 - 8 \Lambda^2} \ ,
\eea
whose solution is $u_1=\pm i 4 \Lambda$ and $u_2= - 6\Lambda^2$.
We thus find exactly four points as expected. They are related by the $\Z_{2n_c - n_f}$ symmetry, which in this case is $\Z_4$.
Again, with these four points, there is a collision with the four singularity ($r=0,1,2$) that departs from $u_1 =0$.
The superpotentials
\beq
\label{superpotentialcoincidencefour}
W(\Phi) = \mu \Tr \left(\frac{\Phi^2}{2} \mp 2 \Lambda  \Phi \right) \ , \qquad  W(\Phi) = \mu \Tr \left(\frac{\Phi^2}{2} \mp 2 i \Lambda  \Phi \right) 
\eeq
select the four coincidence vacua.\footnote{Consider the   singularity $u_2=u_1^2 / 4 -2\Lambda$. If we want to select a generic point on this singularity, then $W = \mu (u_2 - \alpha u_1)= \mu ( u_1^2/4 -2\Lambda - \alpha u_1)$. The vacuum selected will be thus at $u_1 =2 \alpha$ and $u_2 =\alpha^2 -2 \Lambda$.}

For the case $u_1=4\Lambda$, and $u_2=6\Lambda^2$, the roots are all real, and the factorization of the curve is
\beq
y^2 = \frac{1}{4} z^5 (z-2\Lambda)^2 (z-4 \Lambda)
\eeq
Note that $2=4 \left(\cos{\frac{\pi }{4}}\right)^2$. We keep in mind this for the generalization we are going to do in the coming section.


\section{General Case}
\label{General}

We now consider the case of generic $n_c$ and $n_f$. The Seiberg-Witten curve is
\bea
{y}^2 &=& {\cal P}_{(n_c,n_f)}(z) \nonumber \\
&=& \frac{1}{4} \det(z-\phi)^2-  \Lambda^{2n_c-n_f}z^{n_f} \ ,
\eea
where we have defined for convenience the polynomial ${\cal P}_{(n_c,n_f)}(z)$.
The quark  singularities  are labeled by an integer $r$ that runs from $\tnc$ to $[n_f /2]$.
Along these singular sub-manifolds, the curve has a $2r$ zero at the hypermultiplet mass. We consider $m=0$ here for simplicity. The curve is thus factorized as follows
\beq
\label{rvacuafactorization}
{y}^2= z^{2r} {\cal P}_{n_c -r,n_f -2r}(z) \ ,
\eeq
where what remains is the curve for a gauge group $n_c -r$ and $n_f -2r$ flavors.
The adjoint scalar is
\beq
\phi={\rm diag} (0,\dots,0, \phi_{r+1}, \dots, \phi_{n_c} ),
\eeq
where the first $r$ diagonal elements are locked to the mass $m=0$ and the other $n_c -r$  coordinates span the singular sub-manifold of the Coulomb branch.  This manifold is a root of a $r$ non-baryonic Higgs branch.

Let's consider, to begin with, the maximal case $r=[n_f /2]$. Let us restrict also for simplicity to the case $n_f$ even. Along this $n_f /2$ sub-manifold, the SW curve is factorized as follows
\beq
y^2= z^{n_f} {\cal P}_{(n_c - n_f/2,0)}(z) \ .
\eeq
What remains after extracting the $z^{n_f}$ factor is the curve of pure $\U(n_c -n_f/2)$ without matter fields. 
The maximal singularity points for ${\cal P}_{(n_c -n_f/2,0)}(z)$ are given by the solution  
of Douglas and Shenker \cite{Douglas:1995nw}. There are $n_c - n_f/2$ of these maximal singularity points. They arise when the $n_c - n_f/2$ cuts are lined up, and all the roots, apart from two of them, are doubled. The simplest solution is when all the roots are on the real axis. The others are related by a $\exp{\frac{2\pi i k}{2n_c -n_f}}$ transformation.
In the real,  case we have  $\phi = \mathrm{diag} ( \phi_1,\dots, \phi_{n_c -n_f /2}) $ and $\phi_j = 2 \Lambda \cos{ \frac{\pi ( j-1/2)}{n_c - n_f/2}} $. The curve factorization is obtained by using properties of the Chebyshev polynomials:
\bea
\label{factorizationnbb}
{\cal P}_{(n_c -n_f/2,0)}(z) &=& \frac{1}{4} \prod_{j=1}^{n_c - n_f/2}  \left(z - 2 \Lambda \cos{\frac{ \pi (j-1/2) }{n_c - n_f/2}}\right)^2 - \Lambda^{n_c -n_f/2} \nonumber \\
&=& \left( \frac{1}{4} T_{n_c -n_f/2}\left(\frac{z}{2 \Lambda }\right)^2 - 1\right) \Lambda^{n_c -n_f/2} \nonumber \\
&=&   \left(\frac{z^2}{4} -\Lambda^2 \right)  U_{n_c -n_f/2-1}\left(\frac{z}{2 \Lambda}\right)^2   \Lambda^{n_c -n_f/2-2} \ ,
\eea
where $ U_{n_c -n_f/2-1}\left(\frac{z}{2 \Lambda}\right)^2 = \prod_{j=1}^{n_c -n_f/2 -1} \left( \frac{z}{2 \Lambda} -  \cos{  \frac{\pi  j}{n_c -n_f/2}} \right) $. 
In order to factorize the curve, we have used the important identity
\beq
\label{identityP}
T_N ^2 (z) -(z^2 -1)U_{N-1}(z) = 1 \ .
\eeq
The Douglas-Shenker solution provides the exact position of the maximal singularities where all the $\U(1)$ low-energy factors have their own monopole (or dyon) massless.  These discrete vacua are the ones that are selected by the mass perturbation in the superpotential.
The case of $r = n_f /2$ is particularly simple due to the existence of this exact analytic solution. 
The Douglas-Shenker solution will become particularly useful at the end of the section when we shall describe the exact location of the coincidence points.

Now let us describe the low-energy dynamics in a generic $r$ vacuum. We already said that the curve factorizes like Eq.(\ref{rvacuafactorization}) in a sub-manifold of the Coulomb branch of dimension $n_c -r$. An $\SU(r) \times \U(1)_0$ gauge group with $n_f$ flavors in the fundamental representation survives in the low-energy spectrum. The non-Abelian gauge group is infrared free if the condition $r<n_f /2$ is satisfied. It becomes superconformal for the maximal case $r =n_f/2$. The other $n_c -r$ dimensions of the Coulomb branch represent $\U(1)_j$ vector multiplets with $j$ that runs from $1$ to $n_c -r$ . We are then interested in the points of maximal singularity where all the $\U(1)$ gauge groups, except one of them, have their own massless hypermultiplet. In these discrete points, the low-energy physics can thus be summarized in Table \ref{chargesrvacua} containing the gauge groups and corresponding charged hypermultiplets where we have chosen a convenient basis for the cycles in the SW curve so that the charges are all diagonal.   
\TABLE{
\begin{tabular}{c|ccccccccccc}
& &$\SU(r)$ & $\times$ & $\U(1)_0$ & $\times$ &  $\U(1)_1$ & $\times$ &  $\cdots$  & $\U(1)_{n_c-r-1}$ & $\times$ 
& $\U(1)_{n_c -r }$ \\ \hline
${\bf n_f} \times D \phantom{....}$ & &$ \bf r$ & & $1$ &&&             &&       &&      \\
$E_{ 1 \phantom{n_c -r -}}$&        && &  &&        $1$    &&&        &&       \\
$ \vdots \phantom{.....}$& &&&    & &&  & $\ddots$ & &&    \\
$E_{n_c-r-1}$&&             &&&   &&&  & $1$    &  &       \\ \hline
\end{tabular}
\caption{Low-energy particles and charges in a generic $r$ vacuum with $\tnc < r \leq [n_f / 2]$.}
\label{chargesrvacua}
}
It is easy to check that the Higgs
branch emanating from this special vacuum is identical to the baryonic
Higgs branch determined in the classical theory.
These $r$ vacua, where all the particles can be put in a diagonal and local form, are the complete list of the critical points located at $u_1 =0$ that survive after the perturbation by a mass term $\mu \Tr \Phi^2 /2$.
Let us call $A_{\bf r}$ the adjoint chiral superfield of the $\N=2$ $\SU(r)$ vector multiplet, $A_j$ with $j=0, \dots, n_c -r$ the real chiral superfield for the $\U(1)_j$ vector multiplets, and $D,\tD$, $E_j, \tE_j$ the chiral superfields of the matter hypermultiplets. The $\N=2$ low-energy Lagrangian simply follows from the information provided in Table \ref{chargesrvacua}.  After breaking to $\N=1$ with the mass term $\mu \Tr \Phi^2 /2$ in the microscopic theory, the low-energy effective superpotential is
\beq
{\cal W}_{r}^\prime  = \sqrt{2} \left( \tD_i A_{\bf r } D^i + \tD_i A_0  D^i  +\sum_{j=1}^{n_c -r -1} \widetilde{E}_j  A_j E_j - \mu u_2(A_{\bf r},A_0, \dots, A_{n_c -r}) \right) \ .
\eeq
All the matter fields $\tD D$ and $\tE_j E_j$ acquire a condensate due to the vanishing condition for the corresponding $F_{A_{\bf r}}$ and $F_{A_j}$ terms. All the gauge groups, except from the last one $\U(1)_{n_c -r}$, are then Higgsed at an energy scale $\sim \sqrt{\mu \Lambda}$. The theory in the IR thus loses completely the information about the non-Abelian nature of the microscopic theory.

We now come to the minimal case when $r=\tnc$. 
It is better to consider this case separately due to some peculiarities that shall soon be evident. 
The curve in the $r=\tnc$ sub-manifold factorizes like 
\beq
\label{curvebb}
{y}^2= z^{2\widetilde{n}_c} {\cal P}_{(2n_c-n_f,2n_c-n_f)}(z) \ . 
\eeq
What emerges is the curve for gauge group $\U(2n_c -n_f)$ with a number of flavors  $2n_c -n_f$.  There is now a particularly nice solution for the points of maximal singularity. They actually consist of a single vacuum, that is invariant under the $\Z_{2n_c -n_f}$ symmetry of the theory. The location of this point is given by  
\beq
\phi=(0,\dots,0,\Lambda \omega_{2n_c -n_f}, \Lambda \omega_{2n_c - n_f}^2 , \dots, \Lambda \omega_{2n_c - n_f}^{-1} ,\Lambda  ) \ ,
\eeq
where $\omega_{2n_c -n_f}$ is the $(2n_c -n_f)$'th root of unity.
The factorization of the curve is given by the following algebraic steps
\bea
{\cal P}_{(2n_c-n_f,2n_c-n_f)}(z)  &=& \frac{1}{4} \prod_{j=1}^{2n_c -n_f} (z- \Lambda \omega_{2n_c - n_f}^j)^2 - \Lambda^{2n_c-n_f}z^{2 n_c - n_f}    \nonumber \\
&=& \frac{1}{4} \left(z^{2n_c -n_f} + \Lambda^{2n_c - n_f} \right)^2-  \Lambda^{2n_c-n_f}z^{2n_c -n_f} \nonumber\\
&=& \frac{1}{4} \left(z^{2n_c -n_f} - \Lambda^{2n_c - n_f} \right)^2 \ .
\label{factorizationbb}
\eea
\vskip 0.10cm
\noindent
Note the peculiarity that all the roots are now doubled. That is what makes the $\tnc$ vacuum different from the other generic $r$ vacua. Away from $u_1 =0$, the singularity splints into $2n_c -n_f$ different branches. As an example, consider $2n_c -n_f =2$ where two lines depart from the $\tnc$ vacuum as in Figure \ref{plot}.

The $r=\tnc$ maximal critical point, called the root of the baryonic branch in \cite{Argyres:1996eh}\footnote{Since we are now working in $\U(n_c)$ and not $\SU(n_c)$, there is no baryonic branch, only a $\tnc$ non-baryonic Higgs branch.}, is a single point,
invariant under the discrete global $\Z_{2n_c-n_f}$ symmetry of the
theory (the anomaly-free part of the classical $\U(1)$ R-symmetry).
The peculiarity with respect to the previously discussed $r$ vacua is that now there is an extra degeneracy.
The curve given by (\ref{curvebb}) and (\ref{factorizationbb}) has in fact no unpaired roots. This means that now every $\U(1)$ factor in the low energy has its own low-energy massless hypermultiplet. Table \ref{chargesrvacua} must now be supplemented with an additional hypermultiplet.  By
an appropriate choice of basis for the $\U(1)$'s, the charges can be
taken to be as in Table \ref{tnclowenergy}.\footnote{Since we are in $\U(n_c)$ and not $\SU(n_c)$, we can choose a basis so that all the charges are diagonal, even in this maximal singularity case.} 
\TABLE[h]{\begin{tabular}{c|cccccccccc}
& $\SU(\tnc)$ & $\times$ & $\U(1)_0$ & $\times$ &  $\U(1)_1$ & $\times$ & $\cdots$ &   $\U(1)_{2n_c-n_f-1}$
& $\times$ &  $\U(1)_{2n_c -n_f }$ \\ \hline 
${\bf n_f} \times D \phantom{....}$ &$\bf \tnc$ & &  $\phantom{-}1$ & & &            &  &   &&     \\
$E_{1 \phantom{n_c -n_f -1}} $    &   &  &&&      $  \phantom{-}1$  & &   &      &&       \\
 $\vdots \phantom{....} $  && & &&&  & $\ddots$ && &   \\
$E_{2n_c - n_f -1}$ &             & &&&  &&  &$ \phantom{-}1 $&     &       \\
$E_{2n_c - n_f \phantom{-1}}$   &&  &   &  &   & &    & &       & $\phantom{-}1$     \\ \hline
\end{tabular}
\caption{Low-energy particles and charges in the $r=\tnc$ vacuum.}
\label{tnclowenergy}}

Now let us examine the breaking of the effective theory at the $\tnc$ singular point.  In this case, the superpotential is
\bea
{\cal W}_{r=\tnc}^\prime   &=& \sqrt{2}   \Big( \tD_i A_{\rbf} D^i + \tD_i A_0 D^i + \Big. \nonumber \\   &&  \Big.  + \sum_{j=1}^{2n_c -n_f } \tE_j A_j E_i  - \mu u_2(A_{\rbf},A_0, \dots, A_{n_c -r} ) \Big)   \ .
\eea
The important difference with the generic $r$ vacua is that now there is an hypermultiplet $E_j$ for each $\U(1)$ and the dual-quarks $\tD$, $D$ are now relieved from the duty of condensation. The role of $W^{\prime 2}+f$ is now played by two roots in zero.  All the $\U(1)$ factors are then Higgsed and can be integrated out. The low-energy theory is thus a non-Abelian $\SU(\tnc)$ gauge theory with the effective superpotential
\beq
\label{suptnc}
{\cal W}_{r=\tnc}^\prime  = \sqrt{2} \left( \tD_i A_{\rbf} D^i - \frac{\mu}{2} \tr {A_{\rbf}}^2 \right)  \ .
\eeq
$A_{\rbf}$ is also massive and can be integrated out. We are thus left with $\SU(\tnc)$ $\N=1$ super--QCD with $n_f$ flavors.

This \textit{cannot} be the infrared of pure $\N=1$ SQCD.  According to the Seiberg duality, the IR of pure $\N=1$ SQCD is described by $\SU(\tnc)$ gauge theory with $n_f$ flavors plus a meson $M_i^j$ and an opportune superpotential $\tD M D$ \cite{Seiberg:1994pq}. We have the right gauge group and the right dual-quark, but the meson $M_i^j$ is missing. Note that $A_{\rbf}$ cannot be identified with the meson of Seiberg duality. First of all, $A_{\rbf}$ has no flavor charge. Second,  from (\ref{suptnc}) we can see that it becomes massive after the $\mu$ breaking. Thus, it should be integrated out to get the infrared conformal fixed point.

The cases $n_f = n_c+1$ and $n_f=n_c$ require special attention. In these cases, there is no non-Abelian group $\SU(\tnc)$.  For $n_f =n_c +1$, we can still use Table \ref{tnclowenergy} and just delete the $\SU(\tnc)$ column. For $n_f =n_f$, there is no more a flavor charged particle, and we should also delete the $\U(1)_0$ column and the flavored particle ${\bf n_f} \times D$. 

\vskip 0.50cm
\begin{center}
*  *  *
\end{center}
Now we are going to find the general solution for the coincidence vacua. We need to generalize the findings of Section \ref{Example}.  The task seems apparently difficult, but using a  trick, and the help of the Doulgas-Shenker solution, we shall find quite easily the general solution.  First of all, we have to take a look at the example we already found in the previous section and guess from the particular case.
Then we shall prove that the guess is right.

Consider, for the moment, another theory with double the number of colors and flavors 
\beq
N_c =2n_c \ , \qquad N_f =2 n_f \ .
\eeq
Then take the maximal $r=N_f/2$ vacua for this theory. We already discussed at the beginning of the section the exact solution for maximal $r$ vacua, when the number of flavors is even.
The curve is given by
\bea
Y^2 &=& \frac{1}{4} {\rm det} (Z-\Phi) - \LLambda Z^{N_f} \nonumber \\
&=& Z^{N_f} \, {\cal P}_{(N_c -N_f/2,0)}(Z) \ ,
\eea
and for ${\cal P}_{(N_c -N_f/2,0)}(Z) $ we have the $N_c - N_f /2$ Douglas-Shenker solutions, as in (\ref{factorizationnbb}).
\beq
\label{curvebig}
Y^2 =   Z^{N_f} \left(\frac{Z^2}{4} - \LLambda^2 \right) \prod_{j=1}^{N_c -N_f/2 -1} \left( \frac{Z}{2 \LLambda} -  \cos{\frac{\pi  j}{N_c -N_f/2}} \right)^2  \LLambda^{N_c -N_f/2-2} \ .
\eeq
We called $\Phi$ the adjoint scalar field for the theory $(N_c, N_f)$. The Coulomb moduli space has dimension $N_c$ and is parameterized by the coordinates
\beq
U_k = \frac{1}{k} \Tr \Phi^k \ .
\eeq 
Now note that the roots of the curve (\ref{curvebig}) are symmetric if we exchange $j$ with $N_c -N_f/2 -j$ and simultaneously the sign of the roots. We can thus combine the roots with the same modulus and opposite sign and rewrite the curve in the following way, for $N_f/2$  odd
\beq
\label{curvebigodd}
Y^2 = Z^{N_f} \left(\frac{Z^2}{4} -\LLambda^2 \right)  \prod_{j=1}^{N_c/2 -[N_f/4] } \left( \frac{Z^2}{4 \LLambda^2} -  \left( \cos{  \frac{\pi  j}{N_c -N_f/2}}\right)^2  \right)^2  \LLambda^{N_c -N_f/2 -2} \ ,
\eeq
and in the following for $N_f/2$ even
\beq
\label{curvebigeven}
Y^2 = Z^{N_f} \left(\frac{Z^2}{4} -\LLambda^2 \right) 
\prod_{j=1}^{N_c/2 - N_f/4  -1} \left( \frac{Z^2}{4 \LLambda^2} -  \left( \cos{  \frac{\pi  j}{N_c -N_f/2}}\right)^2  \right)^2  Z^2 \LLambda^{N_c -N_f/2 -2}  \ .
\eeq
Note that this looks quite similar to the coincidence points we found in Section \ref{Example}.

To consolidate this guess, note that for the maximal $r=N_f/2$ vacua we can certainly say that the odd part of the moduli space coordinates vanishes
\beq
U_{{\rm odd}}=0 \ .
\eeq
We can thus focus our attention on the sub-moduli space of $U_{\rm even}$ that has exactly the same dimension of the moduli space for the $(n_c,n_f)$ theory.
We finally make the following mapping between the $(N_c,N_f)$ theory and the original $(n_c,n_f)$:
\beq
Z^2=z \ , \qquad \LLambda^2=\Lambda \ .
\eeq
And between the moduli spaces:
\beq
U_{2k}=u_k \ .
\eeq
Now is just a matter of rewriting (\ref{curvebigodd}) and (\ref{curvebigeven}) with the new coordinates, and we get the following factorization of the SW curve
\beq
y^2  = \frac{1}{4}  z^{2[n_f/2] +1} (z - 4 \Lambda ) \prod_{j=1}^{n_c -[n_f/2] -1}  \left(z - 4 \Lambda \left(\cos{\frac{ \pi j }{2n_c -n_f}}\right)^2   \right)^2 \ .
\eeq
This is the coincidence vacuum we were looking for.  This is valid for both $n_f$ even or odd. At the end, everything still follows from the identity (\ref{identityP}).\footnote{The use of the theory $(N_c,N_f)$ has been only a mathematical trick to get the solution for the coincidence vacua passing through Douglas-Shenker. But maybe there is something physical behind this bigger theory. To pursue this idea, certainly a cubic superpotential should be used.}
We thus have the expected solution with $2[n_f/2] +1$ roots in zeros and all the others, except one, doubled.  
The coordinate $\Tr \phi$ of this coincidence vacuum is given by
\bea
 u_1 &=& U_2 \nonumber \\ \nonumber \\
&=&  \sum_{j=1}^{2n_c-n_f} 2 \Lambda \left( \cos{\frac{ \pi (j-1/2) }{2n_c -n_f}} \right)^2  = (2n_c-n_f) \Lambda \ ,
\eea
with the exception $u_1=0$ for $2n_c -n_f=1$. 
The coordinate $\Tr \phi^2/2$ of this coincidence vacuum is given by
\bea
 u_2 &=& U_4 \nonumber \\ \nonumber \\
&=&  \sum_{j=1}^{2n_c-n_f} 4 \Lambda^2 \left( \cos{\frac{ \pi (j-1/2) }{2n_c -n_f}} \right)^4  = \frac{3}{2} (2n_c -n_f) \Lambda^2 \ ,
\eea
with the exceptions $u_2=0,2$ for $2n_c-n_f=1,2$.
The other coincidence vacua are just obtained with a $\Z_{2n_c -n_f} \in \U(1)_R$ transformation.
To select these vacua, we need then to use the following superpotential
\beq
\label{superpotentialcoincidencegeneral}
W(\Phi)=\mu \Tr \left(\frac{\Phi^2}{2} -2 \Lambda e^{\frac{i \, 2\pi k}{2n_c -n_f}} \Phi \right) \ , \qquad k= 1, \dots, 2n_c -n_f \ ,
\eeq
where the index $k$ corresponds to the various coincidence vacua. 
As in the $n_c=2$, $n_f =2$ example, the macroscopic dynamic in these coincidence points is a 
non-local superconformal field theory.


We are also now ready to understand the multiple collision anticipated in Figure \ref{higgs-branches-collision}. At the level $\Tr \phi=0$ of the moduli space, there are multiple discrete $r$ vacua where the curve is factorized similar to Eq.~(\ref{rvacuafactorization}), and the residual part ${\cal P}_{(n_c -r,n_f -2r)}(z)$ is further factorized so that all roots are doubled with the exception of two of them.  These two unpaired roots correspond the roots of $W^{\prime}(z)^2$, split by the quantum effect of the polynomial $f(z)$. We want to start from any one of these $r$ vacua and leave the $u_1 =0$ plane, but remaining in this maximal singularity sub-manifold (that is, all the roots paired except two of them). One complex parameter, $u_1$, can thus be adjusted in order to bring one of the unpaired roots anywhere we want in the complex plane. To have a coincidence, we need one of these unpaired roots to collide with the bunch of zeros $z^{2r}$ representing the quark singularity. So it seems that starting from any one of the $r$ vacua, and tuning the coordinate $u_1$, we can create a different coincidence vacuum. The fact is that they all end up in the same coincidence point, as described by Figure \ref{higgs-branches-collision}. The reason is the following: {\it an odd number of zeros $r^{2h+1}$ is not possible unless $h=[n_f /2]$}. That means that if we start from any $r$ vacuum and we tune $u_1$ to bring an extra root in zero, that automatically brings other $[n_f/2]-r$ couples of roots also in zero. This is exactly what happens in Figure \ref{plot} for $(n_c,n_f)=(2,2)$.

Let us check the previous claim.  In the generic $r$ singularity, the curve factorizes as 
\beq
y^2=z^{2r} \left( \frac{1}{4} P_{n_c-r}(z)^2 - \Lambda^{2n_c-2r}z^{n_f -2r} \right)  \ .
\eeq
It is clear that, if $n_f -2r >1$,  is not possible to bring another unpaired root in zero. Since  $P_{n_c-r}(z)^2$ should also vanish, they always come in pairs. When $n_f -2r =0$ or $n_f -2r =1$, it then  is possible to bring a single unpaired root in zero.

We can understand better the intersections of the $r$ branches by considering only the minimal $\tnc$ and the maximal $[n_f/2]$.  Consider also for simplicity $n_f$ even. We have one $\tnc$ singularity,  that is, the intersection of $2n_c - n_f$ different singularity co-dimension two surfaces at the level $u_1=0$.  Then the maximal $r= n_f /2$ non-baryonic roots, which  are $n_c -n_f /2$ vacua previously described.  Each one of the $n_c -n_f /2$ roots collide with two of the surfaces emanating from the baryonic branch. For example in the $n_c,n_f = 2,2$ case the singularity that departs from the $r=1$ vacuum intersects with the two singularities that depart from the $r=0$ vacuum (Figure \ref{plot}). The same thing happens in the general case. Every singularity that departs from an $n_f/2$ vacuum (there are $n_c -n_f /2$ of them) intersects with two singularities of the $2n_c - n_f$ that depart from the $\tnc$ vacuum.


So we can finally try to summarize and conjecture how the pure $\N=1$ super-QCD emerges from these coincidence points. We can consider the simplest example where $n_f =2n_c -2$. In this case, we have only two vacua at $u_1 =0$: the $\tnc$ vacuum and the $[n_f/2]$ vacuum. The situation is completely analogous to the example $n_c=n_f=2$ discussed in Section \ref{Example}. We can thus refer to Figure \ref{plot} to understand what is happening. The $\tnc$ vacuum has  low-energy dynamics summarized in Table \ref{tnclowenergy} while the $[n_f/2]$ vacuum is summarized in Table \ref{chargesrvacua}. The two non-Abelian multiplets are non-local between them. Let's use $D,\tD$ to denote the $\tnc$ dual-quark and $Q,\tQ$ to denote the $[n_f/2]$ low-energy hypermultiplet. We can suspect that in the $\mu \to \infty$ limit the $Q$ and $\tQ$ condense, forming a mesonic bound state $M=\tQ Q$. This is now local with the dual quarks $D$ and $\tD$. This is probably the way Seiberg duality is obtained in the $\mu \to \infty$ limit.

It is good to add more comments about this last, crucial point. The infrared theory at the coincidence point is a superconformal fixed point where many non-local degrees of freedom are entangled together. Since it is not possible to write an effective local Lagrangian, the study of this theory becomes difficult. Our approach has been to consider the theory at the various $r$ vacua, where a low-energy description is available, and progressively move toward this coincidence point.  In these $r$ vacua, there is a $\N=2$,  $\SU(r) \times \U(1)_0$  gauge theory with a dual-quark in the fundamental of flavor ${\bf n_f}$ and in the fundamental of gauge ${\bf n_f}$. The beta function for the non-Abelian gauge coupling is proportional to $2r - n_f$. This is always infrared free (except for the case the maximal branch $r=n_f/2$ when $n_f$ even). The lowest case $r=\tnc$ and the maximal case $r=[n_f /2]$ are particularly important.  As we already mentioned, the first one provides the $\SU(\tnc)$ and the dual-quarks $\tD$, $D$, which are essential degrees of freedom in the Seiberg duality. The last $r=[n_f/2]$ is also important because it gives the maximal non-baryonic branch $r=[n_f/2]$, which is crucial if we want to recover $\N=1$ SQCD in the $\mu \to \infty$ limit. Note that the beta functions for the $\SU(\tnc)$ and $\SU([n_f/2])$ are, respectively, the lowest and the highest.

When the various $r$ vacua collide in the coincidence point, all the degrees of freedom are merged in this superconformal field theory. We cannot write an explicit theory since they are mutually non-local. But we can nevertheless say that these degrees of freedom are there. The weakest $\SU(\tnc)$ and the strongest $\SU([n_f / 2])$ play an essential role. As we break $\N=2$ with the opportune superpotential (\ref{superpotentialcoincidencegeneral}), we select this coincidence point, and we give mass to all the adjoint scalar fields $A_{{\bf r}}$. The dual-quarks do not condense, and so there is no Higgs effect. In the $\mu \to \infty$, we have to compute again the beta functions that are now $3 r -n_f$. In the maximal case, the beta function is proportional to  $3[n_f/2] - n_f$ and is now strong in the infrared. It is the strongest one among the various $r$. This supports the previous claim that the dual-quarks of the maximal branch condense and provide the meson $M_i^j=\tQ_i Q^j$, essential in the Seiberg dual theory.

For $n_f=n_c+1$ and $n_f = n_c$, there is no non-Abelian group $\SU(\tnc)$. But still the minimal $r$ vacuum ($r=1$ for the first case and $r=0$ for the second) provides an essential ingredient for the phase of low-energy $\N=1$ SQCD: the baryon. In the $n_f=n_c +1$, it is fundamental in flavor (see Table \ref{tnclowenergy} without the $\SU(\tnc)$ column) while for the $n_c =n_f$ case is flavor neutral (see Table \ref{tnclowenergy} without the $\SU(\tnc)$ and $\U(1)_0$ columns). So the minimal $r=1,0$ vacuum provides the baryon while the maximal $r=[n_f/2]$ vacuum provides the meson.

\section{MQCD}
\label{MQCD}


\FIGURE{
\includegraphics[width=20em]{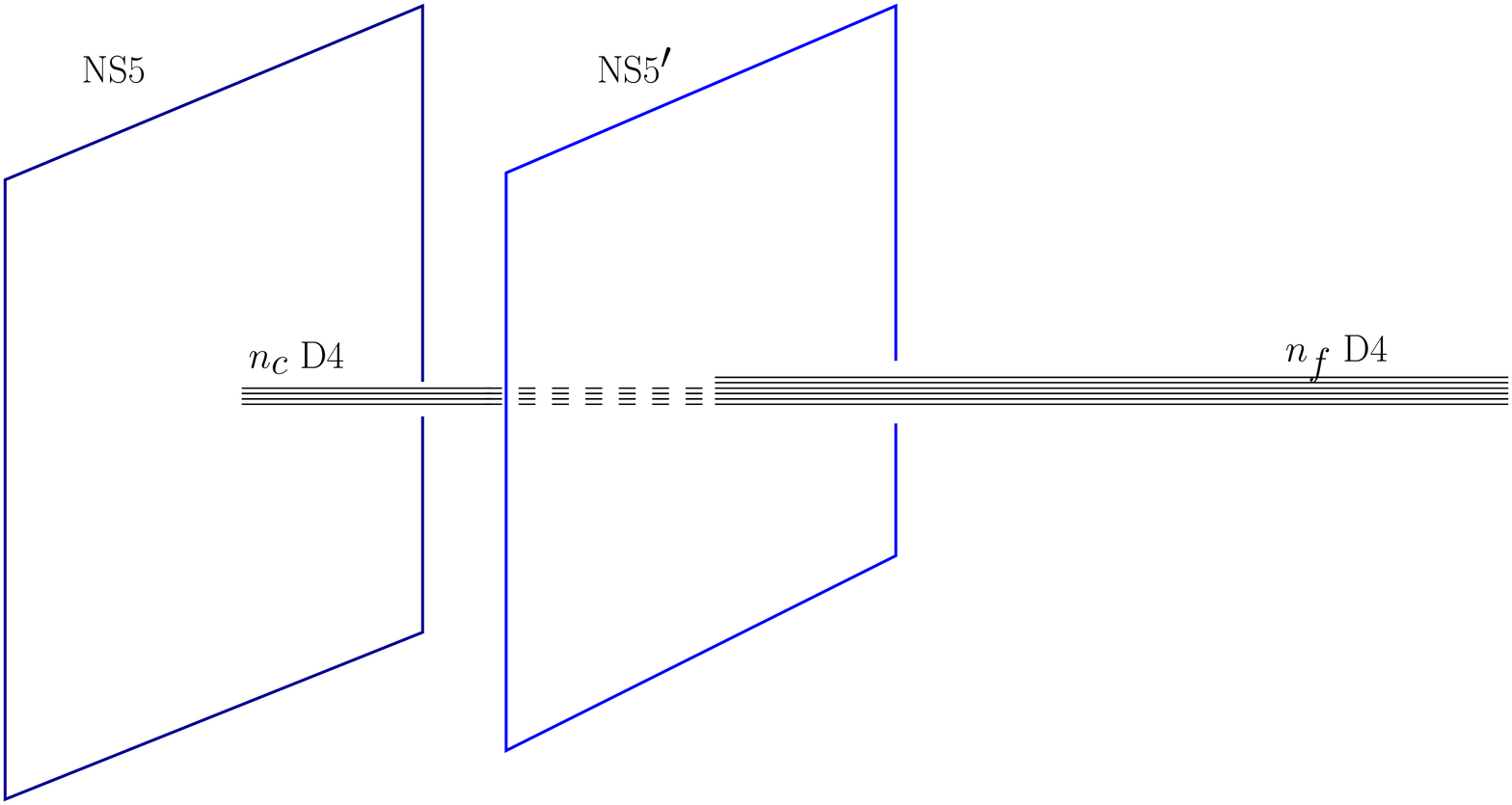}
\caption{Classical brane configuration whose low energy spectrum in $\mathbb{R}^{3,1}$ coincide with super-QCD. }
\label{classical}
}
Super-QCD can be implemented as a low-energy theory of  certain brane setup configurations. A well-known approach consists of taking type IIA string theory and a set of orthogonal NS$5$ and D$4$ branes. Using the common convections, we have the displacement of branes summarized in Table \ref{tablebranesetup} and Figure \ref{classical}.
\TABLE[h!b]{
\begin{tabular}{c|ccccccccccc}
&& $\mathbb{R}^{3,1 = }x^0, \dots, x^3 $ & \,\,\,\, & $v=x^4 + i x^5 $& \,\,\,\, & $x^6 $ &\,\,\,\, & $w = x^7 + i x^8$ &\,\,\,\, & $x^9 $& \\ 
\hline
$\phantom{n_c \,\,\,} \mathrm{NS}5$                  & &$ *$ && $* $&& $0 $   && $0 $     &&  $ 0$      &   \\
$n_c \,\,\, \mathrm{D}4 $       & & $*$ &&$ 0$ && $* $  && $0$      &&  $ 0$      &   \\
$\phantom{n_f \,\,\,} \mathrm{NS}5^\prime$           & & $*$ && $* $&& $L_6$  && $\mu v$  &&   $0 $ &       \\
$n_f \,\,\, \mathrm{D}4  $      & & $* $&& $0 $&& $* $   && $0 $     &&  $ 0 $ & \\
\hline
\end{tabular}\label{tablebranesetup}
\caption{Brane setup in type IIA string theory whose low energy on $\mathbb{R}^{3,1}$ coincide with $\N=2$ and $\N=1$ SQCD.}}
This is just the classical configuration. Taking into account the string coupling effects, the NS$5$ branes are logarithmically bended due to the pulling of the D4 branes. This log-bending corresponds to the running of the coupling constant in the four-dimensional, low-energy action. Other strong coupling effects are related to the junctions between the D$4$ branes and the NS5 branes. A way to resolve these singularities provides a window into the strong coupling of the four-dimensional theory.


One way to study non-perturbative effects is to lift type IIA string theory to M-theory \cite{Witten:1997sc,Giveon:1998sr}. Now the D4 and NS5 branes are all described by the same object, an M$5$ brane. The embedding of the M5 brane is related to the Seiberg-Witten curve and the factorization equation in the $\N=1$ case. We thus obtain a beautiful geometric interpretation of many field theoretical quantities, mostly the chiral and topological ones. The goal for this section is to describe the $r$ vacua and the coincidence vacua in the MQCD framework. The MQCD curve shall provide further evidence for why the coincidence points are so special.

After the M-theory lifting, the internal space is now $Y=\mathbb{R}^6 \times \mathbb{S}^1$, which we parameterize with three complex coordinates $v,\ w$\ and $s$, and one real coordinate $x^7$.  $v$, $w$\ and $x^7$ parameterize the $\mathbb{R}^5 \subset \mathbb{R}^6\times \mathbb{S}^1$ while the complex coordinate $s=x^6+ix^{10}$ parameterizes the remaining $ \mathbb{R} \times \mathbb{S}^1$.  $x^{10}$ is a periodic coordinate parameterizing the M-theory circle. Weakly coupled type IIA string is recovered when the compactification radius is very small.  We define the exponential mapping $t=\exp(s)$, which is valued in $\mathbb{C}^*$.

The brane setup of Figure \ref{classical}, now becomes  a unique M$5$ brane extended along $\mathbb{R}^{3,1}$ times a Riemann surface $\Sigma$ embedded in $Y$. This embedding $\Sigma \subset Y$ contains some crucial information about the quantum field theory of interest.  The surface $\Sigma$ is what finally is related to the Seiberg-Witten data of the original QFT (we have to use the change of coordinates $t=y+P_{n_c}(v)$).  In the $\N=2$ case, the M$5$ brane embedding is the algebraic surface by the equations
\beq
t^2-2P_{n_c}(v)t+\Lambda^{2n_c -n_f} v^{n_f}=0 \ , \label{unot}
\eeq
combined with $w=0$.
The solutions to this equation are 
\beq
\label{duet}
t_{1,2} = P_{n_c}(v) \pm \sqrt{P_{n_c}(v)^2 - 4\Lambda^{2n_c -n_f} v^{n_f} } \ .
\eeq
In the square root, we have exactly the Seiberg-Witten curve. These two branches correspond, asymptotically, to the two NS$5$ branes. The Riemann surface $\Sigma$ is thus a double cover of the $v$ plane.  The solutions at $v \to \infty$ are
\beq
t_1 \sim v^{n_c}\ , \qquad t_2 \sim v^{n_f - n_c}\ .
\eeq
and correspond to the NS$5$ and NS$5^\prime$ branes.
The flavor branes correspond to an expansion around zero. In a generic point of the moduli space, where $P_{n_c}(0) \neq 0$, the flavor brane is attached to the second sheet
\beq
t_1 \sim P_{n_c}(0)\ , \qquad t_2 \sim v^{n_f}\ .
\eeq
These correspond to the $n_f$ flavor D4 branes attached to the NS$5^\prime$ brane.

So far for the $\N=2$ theory. The breaking to $\N=1$ by mean of the superpotential $\Tr W(\Phi)$, correspond in this setting to a deformation of the NS$5^\prime$ brane into the $w$ plane \cite{deBoer:2004he}. From the MQCD perspective, we have to supplement Eq.~(\ref{unot})  with a second equation that provides the information about the embedding in the $w$ plane
\beq
\label{unow}
w^2-2 W^{\prime}(v)w-f(v)=0 \ ,\label{wcond}
\eeq
where the  polynomial $f(v)$  captures the quantum corrections to the superpotential. The two branches of the solutions are
 \beq
 \label{duew}
 w_{1,2} =  W^{\prime}(v) \pm \sqrt{W^{\prime}(v)^2 + f(v) }\ .
 \eeq
Classically, without $f(v)$, we have $w=0$ that corresponds to the NS$5$ brane and $w=2 W^{\prime}(v)$ that corresponds to the NS$5^\prime$ brane deformation in the $w$ plane.


Without the superpotential the $\N=2$, curve (\ref{unot}) enjoys a moduli space of the solutions parameterized by the $n_c$ coefficients contained in the polynomial $P_{n_c}(v)$. But things are more complicated when a superpotential is introduced and we have to deal also with Eq.~(\ref{unow}).
The equations (\ref{duet}) and (\ref{duew}) are both two branches that cover the $v$ plane. Passages from one branch to another are determined by what is inside the square root: the Seiberg-Witten curve in the first case and the $\N=1$ curve in the second.  For a generic value of $P_{n_c}$, we have a total of four branches connected together. So nothing that resembles the classical picture.
Only for a particular discrete set of solutions, exactly when the factorization (\ref{curveandfactorization}) is satisfied,  can  we separate the four branches into two disconnected parts: $t_1, w_1$ together with $t_2, w_2$,  and $t_1, w_2 $ together with $t_2, w_1$. 
And this is  the MQCD explanation for the $\N=1$ factorization of the Seiberg-Witten curve.


For the generic $r$ vacuum, the curve is given by
\[
t^2-2T_{n_c -n_f/2}\left(\frac{v^2}{2}\right)^2 v^{n_f /2} t+\Lambda^{2n_c -n_f} v^{n_f} = 0 \ , 
\]
\beq
 w^2-2 \mu v w- 1  =  0 \ .
\eeq
Flavor branes are separated into two groups of $r$ and $n_f - r$ units (see Figure \ref{nbbc}). These two correspond to two different spikes of the M5 brane separated in the $w$ plane by a distance $\sqrt{W^{\prime}(m)^2 + f(m)}$ (remember the two solutions in Eq.~(\ref{duew})). This in fact corresponds to the dual-quark condensate (\ref{dualcondensate}).
\DOUBLEFIGURE{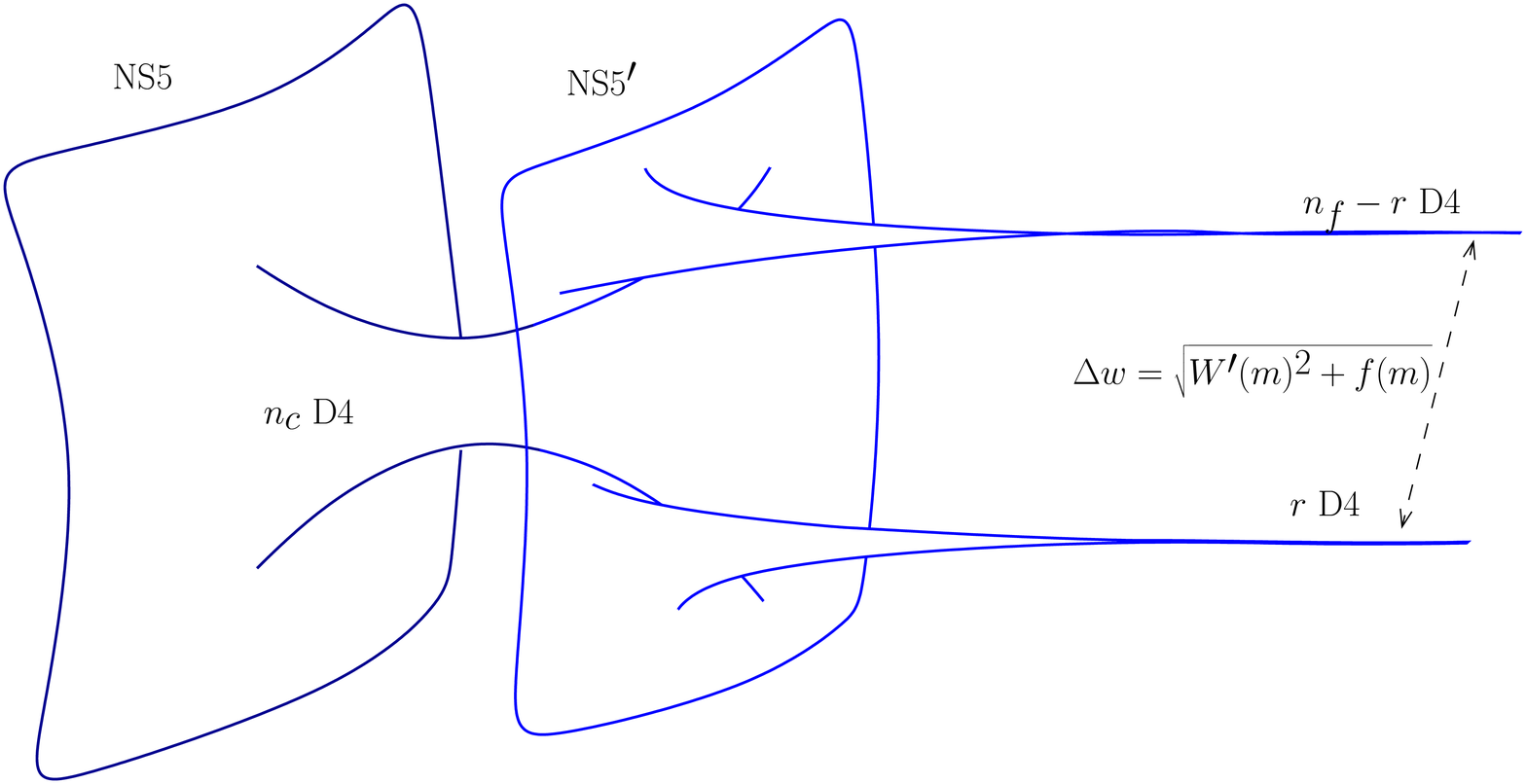,width=22.5em}{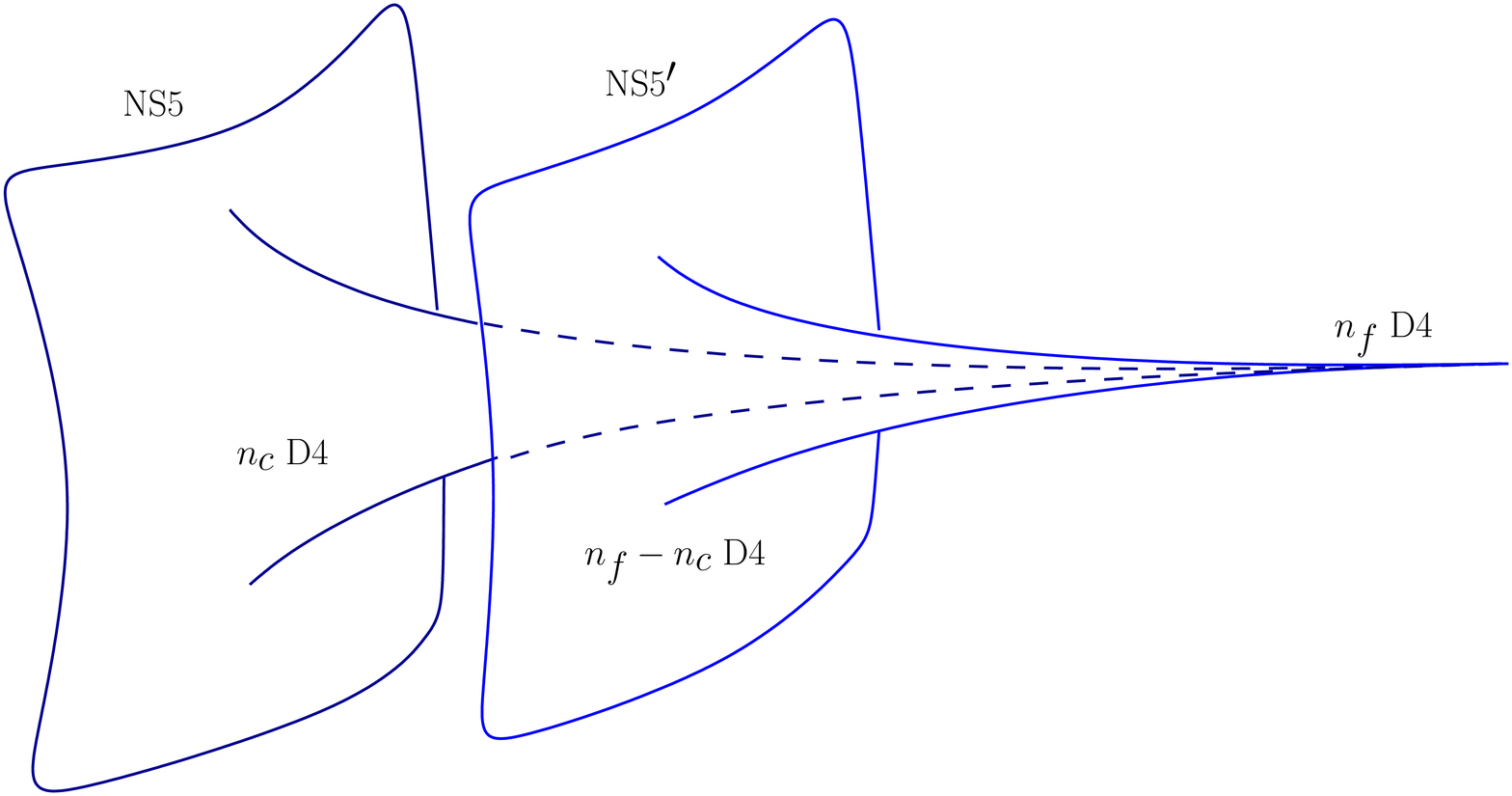,width=21.5em }{ \label{nbbc} MQCD curve corresponding to a generic $r$ vacuum.}{\label{bbc} MQCD curve for the specific case $r=\tnc$. }
The $r=\tnc$ branch is an exception. Now the curve is given by
\[
t^2-2\left( z^{2n_c -n_f} + \Lambda^{2n_c - n_f}\right)  v^{\tnc}  t+\Lambda^{2n_c -n_f} v^{n_f} = 0 \ ,
\]
\beq
 w^2-2 \mu v w  =  0 \ .
\eeq
In this case, there is an extra massless particle, and the MQCD becomes divided into two distinct pieces. The two planes meet only at $x_6 \to \infty $. The reason can be seen from the factorization of the SW curve (\ref{factorizationbb}). All the roots are now paired and positioned at $\Lambda \omega_{2n_c -n_f}^j$. The cuts are in a closed polygonal shape with $2n_c-n_f$ sides; they topologically separate the interior and the exterior of the polygon in the complex $v$ plane.  
A spike corresponding to $n_f -n_c$ D$4$ branes departs from the NS$5^\prime$ branes and joins at $x_6 \to \infty$ with a spike of $n_c$ D$4$ branes departed from the left NS$5$ branes (see Figure \ref{bbc}). There is no asymptotic separation in the $w$ plane and that means no condensation of the dual quark. Note  also that the topology of the curve is different. It consists of two disconnected pieces that join only at $x_6 \to \infty$ (see Figure \ref{bbc}).

\FIGURE[h]{
\includegraphics[width=32em]{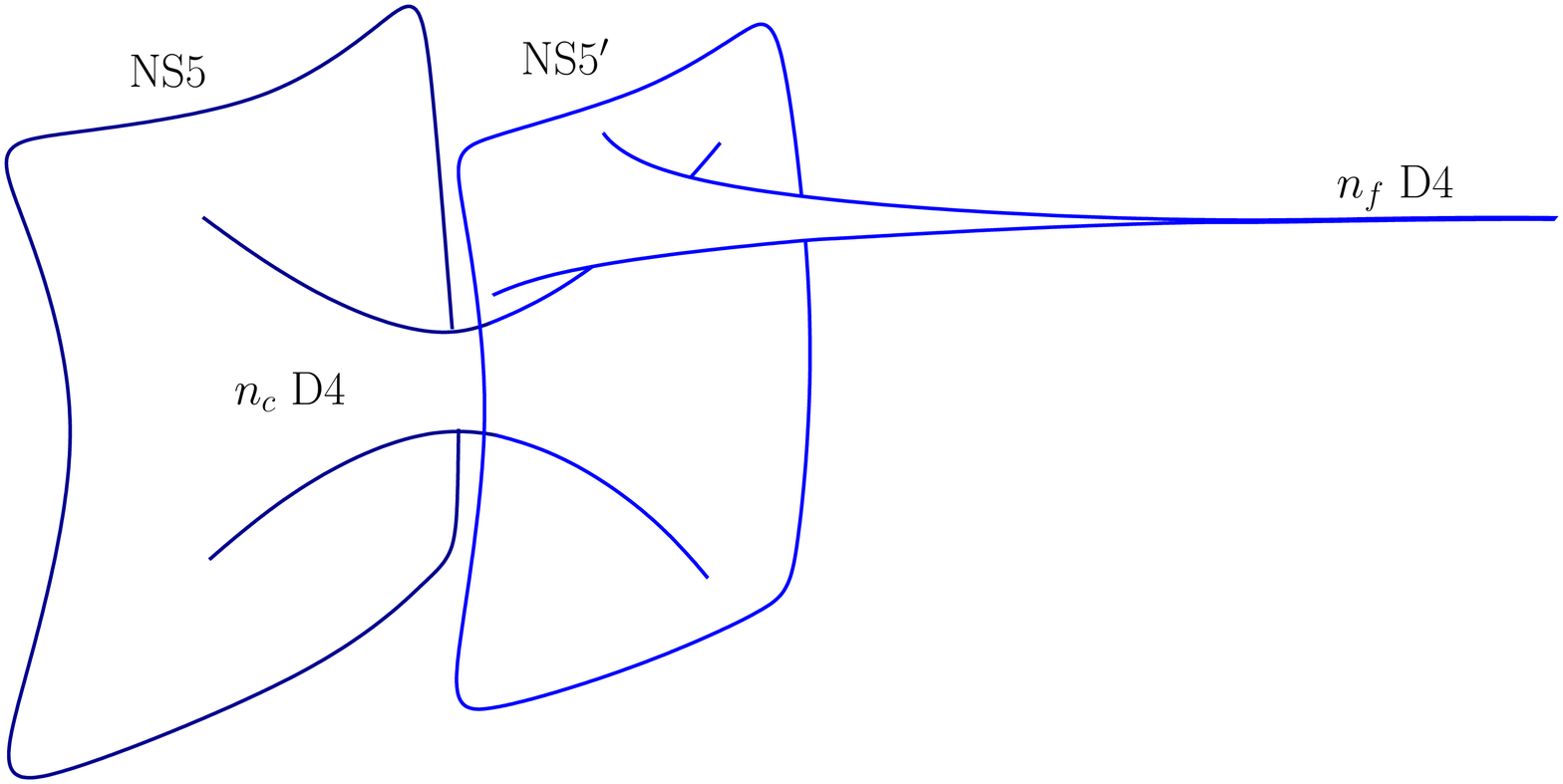}
\caption{MQCD curve for a coincidence point.}
\label{coincidencepointsMQCD}
}
We have thus seen that none of the singularities at $\Tr \phi =0$ have the right MQCD curve to describe $\N=1$ SQCD at $\mu \to \infty$. In all cases, there is a mass gap and nothing in the low energy. In the baryonic root, the quarks have zero expectation value, but the topology of the curve is changed. 
The coincidence points are instead very special. The condensate of the quark is zero, and there is no change of topology. The curve is given by
\[
t^2-2 v^{[n_f /2]} \prod_{j=1}^{n_c -[n_f/2]} \left(z - 4 \Lambda \left(\cos{ \frac{\pi (j-1/2)}{2n_c -n_f}}\right)^2\right)^2 t+\Lambda^{2n_c -n_f} v^{n_f} = 0 \ ,\]
\beq
w^2-2 \mu v w- v   = 0 \ . 
\eeq
Good things happen in this circumstance. First of all, the flavor branes correspond to a unique spike of the M$5$ brane. There is no asymptotic separation in the $w$ plane and that means no condensation. Furthermore, despite what happened for the $\tnc$ vacuum, the Riemann surface preserves its topology. The curve is still a double cover of the $v$ plane (Figure \ref{coincidencepointsMQCD}).



\section{Conclusion}
\label{conclusion}
In the paper, we addressed a problem of extended supersymmetry breaking, from $\N=2$ to $\N=1$. Generically, this is achieved by giving a mass term to the adjoint scalar field $\phi$ of the $\N=2$ gauge supermultiplet. Classically, this works perfectly fine, with or without matter hypermultiplets. When quarks fields are present, some of the flat directions of the $\N=1$ theory are already present in the $\N=2$ theory. Others are recovered as pseudo-moduli with mass proportional to $\propto 1/\mu$, that become massless in the $\mu \to \infty$ limit.

In quantum theory, there are new subtleties that spoil this breaking pattern. First of all, the notion of the origin of the moduli space does not hold anymore. What was the classical origin of the moduli space is in some sense split into various vacua labeled by an integer $r$. A mass term for the adjoint scalar field $\phi$ selects any one of these vacua, but none of them, in the $\mu \to \infty$ limit, flows exactly to pure $\N=1$ super-QCD.

\FIGURE{
\includegraphics[width=21em]{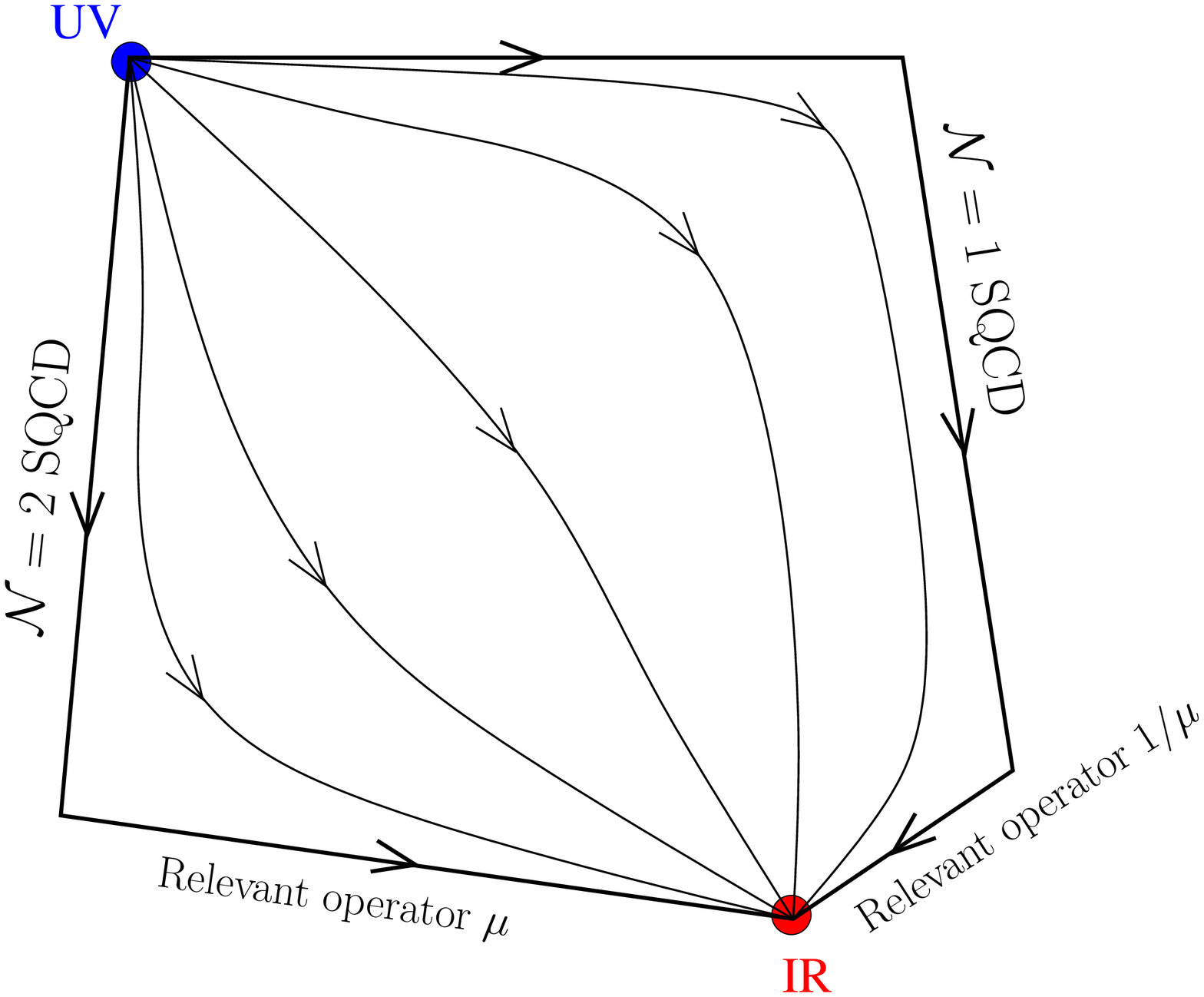}
\caption{RG flow for various values of $\mu$. The five corners are fixed points of the RG flow.}
\label{flow-general}
}
The reason is quantum mechanical and must be traced back to the operator that is generated after integrating out the field $\phi$. This operator
\beq
\label{theoperator}
W =   \frac{1}{{\mu \sqrt{2} }} \ \tQ Q \  \tQ Q \ ,
\eeq
looks apparently harmless. Since it is proportional to $1/\mu$, we would be tempted to conclude that its effects become negligible in the $\mu \to \infty$ limit. But quantum mechanically it leads to an operator of dimension $d$ generally smaller than three. In $\N=1$ SQCD, the dimension of the meson operator is $ D(\tQ Q)= 3 R^{\prime}(\tQ Q)/2 = 3 (n_f -n_c)/n_f$.  So (\ref{theoperator}) leads generally to an operator of dimension 
\beq
D(\tQ Q \  \tQ Q) = 6\frac{n_f-n_c}{n_c} \ ,
\eeq
for very large $\mu$. For $n_f =2n_c$ this is exactly marginal.\footnote{Note the non-triviality of that. The $\N=1$  is in general not sensitive to the equality $n_f=2n_c$ which is instead very important for the $\N=2$ theory.}  It is thus a \textit{relevant} operator, and it grows as $(\epsilon/\epsilon_0)^{d-3}$ as the energy scale of the renormalization group flow $\epsilon$ goes to zero.
No matter how small  the coefficient in front of it is, it will always blow up for sufficiently small energies of the RG flow and force the theory to flow to another IR fixed point. Most of the time, like in the cases $\tnc < r \leq [n_f / 2]$,  the theory flows to nothing, and there is a mass gap for the gauge degrees of freedom. There is a special case, the minimal value $r=\tnc$, where the theory flows to $\SU(\tnc)$ with $n_f$ dual-quarks $\tD, D$. Although very near to what we should expect in pure $\N=1$ SQCD, still there is a missing piece: the meson $M_i^j$ and its superpotential interaction with $\tD, D$. We can interpret this missing piece as damage still caused by the relevant operator (\ref{theoperator}). We can thus conclude that for generic  $\SU(n_c)$ SQCD there is no possibility to flow exactly from $\N=2$ to $\N=1$ (Figure \ref{flow-general}).

But the classical analysis comes back as a source of inspiration. As we saw from the prototype model, super-QED with hypermultiplets of mass $m$ and superpotential $W(\phi)$, and the classical theories have a large spectrum of vacua, and a few of them, in general, flow to pure $\N=1$ super-QED as the superpotential goes to infinity. There are Coulomb vacua where $\phi$ is equal to the root of the superpotential and the quarks are massive. There are gauge-flavor locked vacua where $\phi$ is locked to some quark mass and the gauge group is broken by the quark condensate $W^{\prime}(m)$. Finally, there are the ``coincidence'' vacua that arise under the very particular circumstance in which a root $a$  of $W^\prime$ precisely coincides with a hypermultiplet mass $m$. Choosing $\phi = m =a$, we get a particular vacuum in which the quarks are massless, but they do not condense. These are the right vacua in which to flow from $\N=2$ to $\N=1$ super-QED with massless quarks.

Returning to the non-Abelian case, we thus decided to extend the space of interest from $\SU(n_c)$ to $\U(n_c)$. $\U(n_c)$ $\N=2$ super-QCD has, with respect to $\SU(n_c)$, one dimension more in the Coulomb moduli space given by the coordinate $u_1 = \Tr \phi$. This extra dimension is crucial. The quantum effects, as already said, erase the notion of the origin of the moduli space and split it into various $r$ vacua (in the $\Tr \phi=0$ section). None of these vacua can be considered as the right generalization of the classical notion of coincidence vacua. None flows to $\N=1$ super-QCD after the mass breaking term is sent to infinity. But extending the search to the $u_1$ dimension we find a nice surprise. We find $2n_c -n_f$ points that, by all rights, can be considered as the quantum generalization of the classical notion of coincidence vacua.
With a suitable superpotential
\beq
W(\Phi)=\mu \Trnc \left(\frac{\Phi^2}{2} - 2\Lambda e^{\frac{i \, 2\pi k}{2n_c -n_f}} \Phi \right) \ , \qquad k= 1, \dots, 2n_c -n_f \ ,
\eeq
we can select any one of these vacua. And each one is a good coincidence point.

The \textit{statement} is thus that these coincidence points are the right ones to flow from $\N=2$ super-QCD to $\N=1$ super-QCD. They are the points in which the effects of the relevant operator (\ref{theoperator}) are minimized and the theory flows as close as possible to $\N=1$ super-QCD.  Let us summarize the arguments we presented in the paper.
\begin{itemize}
\item  The  formula for the quark condensate, $\tq q \propto \sqrt{W^{\prime}(m)^2 + f(m)} $, is the starting point in the search for the generalization of the classical notion of coincidence. Quantum mechanically, the roots of $W^{\prime}(z)^2$ are in general split by the polynomial $f(z)$. For the quark condensate to vanish, we need to send one of these roots near the bunch of zeros at the mass $m$. This guiding principle gives us the location of the coincidence points and the factorization of the Seiberg-Witten curve in these vacua.
\item  Having done this, we achieve also another, a priori not required, result. These coincidence points are located at the intersection between singularities that depart from all the $r$ vacua, from $\tnc$ to $[n_f/2]$ (the collision of Figure \ref{higgs-branches-collision}). We thus recover the notion of the origin of the moduli space that in the $\Tr \phi=0$ section of the moduli space was lost.  The Higgs branch emanating from these vacua is equal to the maximal non-baryonic branch for $r=[n_f/2]$.  But the theory at the root of the branch is much richer than the original $r=[n_f / 2]$ vacuum at $\Tr \phi = 0$. It is a non-local strongly interacting theory, a particular kind of Argyres-Douglas singularity. 
\item  The MQCD approach provides an interesting point of view, giving some geometric intuition about what is going on. The classical brane setup consists of NS$5$ branes and perpendicular D$4$ branes. Some quantum effects, such as the Seiberg-Witten curve and the factorization due to the superpotential, can be analyzed by lifting to M-theory where the branes are described by a single M$5$-brane with a proper embedding in the $v,t,w$ space. In ordinary $r$ vacua, the flavor D$4$-branes are divided into two sets, $r$ and $n_f -r$, separated by a $w$ distance of $\sqrt{W^{\prime}(m)^2 + f(m)}$. This is the signal of quark condensation. For the particular case $r=\tnc$, the flavor branes are not separated in the $w$ plane (the quarks do not condense). But in this particular case the M$5$ brane gets divided into two separated curves that meet only at $v \to +\infty$. Only for the coincidence vacua we have are the quarks D$4$-branes non-separated and the topology of the curve does not change.  This is certainly what is closer to the classical realization of a rotation of the NS$5^\prime$ brane while keeping the D$4$ gauge and flavor branes all allineated.  
\item  We said that the vacua at $u_1 =0$ are not the right coincidence ones because of the relevant operator (\ref{theoperator}). In $\mu \to \infty$, the coefficient in front of the operator  goes to zero, but the operator itself always goes to infinity for sufficiently small energies of the RG flow. This is a relevant perturbation that makes the theory flow down from the pure $\N=1$ super-QCD infrared fixed point. In general, we cannot predict where the theory will end, but one thing we can certainly say: {\it it is something smaller than the pure $\N=1$ vacuum}. We should not underestimate this piece of information. The generic $r$ vacua are trivial examples, since the theory flows to nothing. The $r=\tnc$ is an interesting case. From $\N=1$ $\SU(\tnc)$ with $n_f$ dual-quarks $\tD$, $D$ and the meson $M_i^j$ (the IR of pure $\N=1$ SQCD), we flow to the same theory, just without the meson. 

The coincidence points, as part of the $r=\tnc$ singularity, contain the degrees of freedom of the gauge group $\SU(\tnc)$ and the dual-quark $\tD$, $D$. Since they are also part of the other $r$ singularities, they contain many more degrees of freedom.  It is hard to imagine that there is something different where the theory could flow, lower than pure $\N=1$ and higher than the one at the $r=\tnc$ vacuum.  What is missing is just the meson, and, as we saw, the coincidence vacua are certainly capable of providing it.
\end{itemize}

\vskip 0.10cm
\noindent
\TABLE{
\begin{tabular}{cccc}
\hline
&$n_f < n_c$ & \, & $\mu \Lambda_{\snd} \longrightarrow \infty \phantom{ 0 \Lambda_{\snu}^2}$ \\
&$n_f = n_c$ & \, & $\mu \Lambda_{\snd} \longrightarrow  \Lambda_{\snu}^2 \phantom{ 0 \infty} $ \\
&$n_f > n_c$ & \, & $\mu \Lambda_{\snd} \longrightarrow 0 \phantom{ \infty \Lambda_{\snu}^2} $ \\ \hline
\end{tabular}
\caption{Scaling of $\mu \Lambda_{\snd}$ as $\mu \to \infty$.}
\label{scalingmulambda}
}
The conclusion is that these particular coincidence points should be considered as the quantum analog of coincidence between the hypermultiplet mass and the root of the $W^\prime$. 
Although the quark condensate vanishes, it is plausible that the Higgs moduli space will be modified by the $\mu \Lambda_{\snd}$ term. In Table \ref{scalingmulambda}, we have the $\mu \Lambda_{\snd}$ condensate scales as $\mu$ is sent to infinity (remember the relationship (\ref{relationofscales}) between the two scales). This is probably related to the fact that the moduli space for $\N=1$ SQCD is quantum modified for $n_f=n_c$ and not modified from the classical one for $n_f > n_c$.

Note that, at the contrary of color-flavor locked vacua, coincidence vacua can also exist  for $n_f < n_c$. Since the color branes are locked to a root of $W^\prime$, there is no lower bound on the number of flavors we can attach to the color branes. Although we have focused our attention on $n_f \geq n_c$, many things go unchanged for $n_f$ smaller. In particular, there are still $2n_c -n_f$ coincidence points in the moduli space and their position and curve are still described by the findings of Section \ref{General}. It is known that $\N=1$ SQCD for $n_f <n_c$ has an instanton generated run-away potential. This could be probably related to the scaling of $\mu \Lambda_{\snd}$ of Table \ref{scalingmulambda}.

We are left with challenging questions for the future.  One question is what happens to the operator (\ref{theoperator}) in the coincidence vacua. Another question regards the $\mu$ transition from $0$ to $\infty$.
The superconformal field theory at the coincidence vacua is like a boiling soup, with many non-local degrees of freedom. We have conjectured how the dual-quark and the meson of the Seiberg duality emerge out of it in the $\mu \to \infty$ limit. Certainly, a more detailed understanding of this transition is needed.
In particular, we do not know if, in the IR fixed point, this transition is sharp or if it is a marginal deformation from the superconformal $\N=2$ to the infrared of the Seiberg duality. The completion of this program should eventually be considered the field theoretical proof of the Seiberg duality that was initiated in \cite{Argyres:1996eh}.


\acknowledgments
I want to thank M.~Shifman and A.~Yung for useful discussions about susy-QCD, and in particular the  heterotic vortex problem. I am grateful to M.~Shifman for his help and support. I want to thank K.~Konishi for discussions in the past regarding the susy-QCD moduli space and the APS paper.  I want also to thank various collaborators who helped me in the past in the study of related problems: R.~Auzzi, J.~Evslin and M.~Matone. I want to thank Ki-Myeong Lee and people at KIAS for their hospitality in early June 2008. I want also to thank A.~Vainshtein for recent discussions.  This work is supported by DOE grant  DE-FG02-94ER40823.


\begin{thebibliography}{99}
{\small \itemsep -2pt }







\bibitem{Argyres:1996eh}
  P.~C.~Argyres, M.~R.~Plesser and N.~Seiberg,
  ``The Moduli Space of N=2 SUSY {QCD} and Duality in N=1 SUSY {QCD},''
  Nucl.\ Phys.\  B {\bf 471}, 159 (1996)
  [arXiv:hep-th/9603042].
   



  \bibitem{Bolognesi:2004da}
S.~Bolognesi,
  ``The holomorphic tension of vortices,''
  JHEP {\bf 0501}, 044 (2005)
  [arXiv:hep-th/0411075].
  
  S.~Bolognesi,
  ``The holomorphic tension of nonabelian vortices,''
  Nucl.\ Phys.\  B {\bf 719}, 67 (2005)
  [arXiv:hep-th/0412241].

  
  
  
\bibitem{Shifman:2005st}
  M.~Shifman and A.~Yung,
  ``Non-abelian flux tubes in SQCD: Supersizing world-sheet supersymmetry,''
  Phys.\ Rev.\  D {\bf 72}, 085017 (2005)
  [arXiv:hep-th/0501211].
   

\bibitem{Edalati:2007vk}
  M.~Edalati and D.~Tong,
  ``Heterotic vortex strings,''
  JHEP {\bf 0705}, 005 (2007)
  [arXiv:hep-th/0703045].
  
   D.~Tong,
  ``The quantum dynamics of heterotic vortex strings,''
  JHEP {\bf 0709}, 022 (2007)
  [arXiv:hep-th/0703235].


\bibitem{Shifman:2008wv}
  M.~Shifman and A.~Yung,
  ``Heterotic Flux Tubes in N=2 SQCD with N=1 Preserving Deformations,''
  arXiv:0803.0158 [hep-th].

    M.~Shifman and A.~Yung,
  ``Large-N Solution of the Heterotic N=(0,2) Two-Dimensional CP(N-1) Model,''
  arXiv:0803.0698 [hep-th].
  
  
\bibitem{Auzzi:2004yg}
  R.~Auzzi, S.~Bolognesi and J.~Evslin,
  ``Monopoles can be confined by 0, 1 or 2 vortices,''
  JHEP {\bf 0502}, 046 (2005)
  [arXiv:hep-th/0411074].
 
  
  S.~Bolognesi and J.~Evslin,
  ``Stable vs unstable vortices in SQCD,''
  JHEP {\bf 0603}, 023 (2006)
  [arXiv:hep-th/0506174].

  

  
  \bibitem{Intriligator:1995au}
  K.~A.~Intriligator and N.~Seiberg,
  ``Lectures on supersymmetric gauge theories and electric-magnetic  duality,''
  Nucl.\ Phys.\ Proc.\ Suppl.\  {\bf 45BC}, 1 (1996)
  [arXiv:hep-th/9509066].
  
  
  \bibitem{Seiberg:1994aj}
  N.~Seiberg and E.~Witten,
  ``Electric - magnetic duality, monopole condensation, and confinement in N=2
  supersymmetric Yang-Mills theory,''
  Nucl.\ Phys.\  B {\bf 426}, 19 (1994)
  [Erratum-ibid.\  B {\bf 430}, 485 (1994)]
  [arXiv:hep-th/9407087].
  
  
  N.~Seiberg and E.~Witten,
  ``Monopoles, duality and chiral symmetry breaking in N=2 supersymmetric
  QCD,''
  Nucl.\ Phys.\  B {\bf 431}, 484 (1994)
  [arXiv:hep-th/9408099].

  
  
\bibitem{Cachazo:2002ry}
F.~Cachazo, M.~R.~Douglas, N.~Seiberg and E.~Witten,
  ``Chiral rings and anomalies in supersymmetric gauge theory,''
  JHEP {\bf 0212}, 071 (2002)
  [arXiv:hep-th/0211170].
  
 F.~Cachazo, N.~Seiberg and E.~Witten,
  ``Phases of N = 1 supersymmetric gauge theories and matrices,''
  JHEP {\bf 0302}, 042 (2003)
  [arXiv:hep-th/0301006].


  

  
\bibitem{Cachazo:2003yc}
   F.~Cachazo, N.~Seiberg and E.~Witten,
  ``Chiral rings and phases of supersymmetric gauge theories,''
  JHEP {\bf 0304}, 018 (2003)
  [arXiv:hep-th/0303207].
  
  

   \bibitem{Argyres:1995jj}
  P.~C.~Argyres and M.~R.~Douglas,
  ``New phenomena in SU(3) supersymmetric gauge theory,''
  Nucl.\ Phys.\  B {\bf 448}, 93 (1995)
  [arXiv:hep-th/9505062].

 
\bibitem{Argyres:1995xn}
  P.~C.~Argyres, M.~Ronen Plesser, N.~Seiberg and E.~Witten,
  ``New N=2 Superconformal Field Theories in Four Dimensions,''
  Nucl.\ Phys.\  B {\bf 461}, 71 (1996)
  [arXiv:hep-th/9511154].
  
  
 \bibitem{Eguchi:1996vu}
  T.~Eguchi, K.~Hori, K.~Ito and S.~K.~Yang,
  ``Study of $N=2$ Superconformal Field Theories in $4$ Dimensions,''
  Nucl.\ Phys.\  B {\bf 471}, 430 (1996)
  [arXiv:hep-th/9603002].

  
  \bibitem{Gorsky:2000ej}
  A.~Gorsky, A.~I.~Vainshtein and A.~Yung,
  ``Deconfinement at the Argyres-Douglas point in SU(2) gauge theory with
  broken N = 2 supersymmetry,''
  Nucl.\ Phys.\  B {\bf 584}, 197 (2000)
  [arXiv:hep-th/0004087].
  
  

\bibitem{Douglas:1995nw}
  M.~R.~Douglas and S.~H.~Shenker,
  ``Dynamics of SU(N) supersymmetric gauge theory,''
  Nucl.\ Phys.\  B {\bf 447}, 271 (1995)
  [arXiv:hep-th/9503163].
  
  
\bibitem{Seiberg:1994pq}
  N.~Seiberg,
  ``Electric - magnetic duality in supersymmetric nonAbelian gauge theories,''
  Nucl.\ Phys.\  B {\bf 435}, 129 (1995)
  [arXiv:hep-th/9411149].  
 
 
  N.~Seiberg,
  ``Exact Results On The Space Of Vacua Of Four-Dimensional Susy Gauge Theories,''
  Phys.\ Rev.\  D {\bf 49}, 6857 (1994)
  [arXiv:hep-th/9402044].  
  
  \bibitem{Pisa}
     R.~Auzzi, S.~Bolognesi, J.~Evslin, K.~Konishi and A.~Yung,
  ``Nonabelian superconductors: Vortices and confinement in N = 2 SQCD,''
  Nucl.\ Phys.\  B {\bf 673}, 187 (2003)
  [arXiv:hep-th/0307287].
  
  S.~Bolognesi and K.~Konishi,
  ``Non-Abelian magnetic monopoles and dynamics of confinement,''
  Nucl.\ Phys.\  B {\bf 645}, 337 (2002)
  [arXiv:hep-th/0207161].
  
  G.~Carlino, K.~Konishi and H.~Murayama,
  ``Dynamical symmetry breaking in supersymmetric SU(n(c)) and USp(2n(c))
  gauge theories,''
  Nucl.\ Phys.\  B {\bf 590}, 37 (2000)
  [arXiv:hep-th/0005076].
  
  
   

  
  
\bibitem{Witten:1997sc}
  E.~Witten,
  ``Solutions of four-dimensional field theories via M-theory,''
  Nucl.\ Phys.\  B {\bf 500}, 3 (1997)
  [arXiv:hep-th/9703166].

  E.~Witten,
  ``Branes and the dynamics of {QCD},''
  Nucl.\ Phys.\  B {\bf 507}, 658 (1997)
  [arXiv:hep-th/9706109].
  
  
  
\bibitem{Giveon:1998sr}
  A.~Giveon and D.~Kutasov,
  ``Brane dynamics and gauge theory,''
  Rev.\ Mod.\ Phys.\  {\bf 71}, 983 (1999)
  [arXiv:hep-th/9802067].


  
  
\bibitem{deBoer:2004he}
  J.~de Boer and S.~de Haro,
  ``The off-shell M5-brane and non-perturbative gauge theory,''
  Nucl.\ Phys.\  B {\bf 696}, 174 (2004)
  [arXiv:hep-th/0403035].
  
   J.~de Boer and Y.~Oz,
  ``Monopole condensation and confining phase of N = 1 gauge theories via M-theory fivebrane,''
  Nucl.\ Phys.\  B {\bf 511}, 155 (1998)
  [arXiv:hep-th/9708044].
  
  
  
  


  

\end{thebibliography}
\end{document}